\documentclass[12pt,a4paper]{article}
\usepackage{amssymb,amsmath,amsthm,euscript}
\usepackage{epic}
\usepackage{eepic}

\newcommand*{\hm}[1]{#1\nobreak\discretionary{}%
            {\hbox{$\mathsurround=0pt #1$}}{}}
\newtheorem{theorem}{Theorem}
\newtheorem{prop}{Proposition}
\newtheorem{lemma}{Lemma}
\newcommand{\ch}{\mathop{\mathrm{ch}}\nolimits}
\newcommand{\sh}{\mathop{\mathrm{sh}}\nolimits}
\newcommand{\dn}{\mathop{\mathrm{dn}}\nolimits}
\newcommand{\cn}{\mathop{\mathrm{cn}}\nolimits}
\newcommand{\sn}{\mathop{\mathrm{sn}}\nolimits}
\newcommand{\End}{\mathop{\mathrm{End}}\nolimits}
\newcommand{\tr}{\mathop{\mathrm{tr}}\nolimits}
\newcommand{\sign}{\mathop{\mathrm{sign}}\nolimits}
\numberwithin{equation}{section}
\renewcommand{\Im}{\mathop{\mathrm{Im}}\nolimits}
\renewcommand{\Re}{\mathop{\mathrm{Re}}\nolimits}

\author{A.\,Silantyev \footnote{E-mail: silant@thsun1.jinr.ru, silant@tonton.univ-angers.fr} \bigskip\\
{\normalsize \itshape Bogoliubov Laboratory of Theoretical Physics, JINR } \\
{\normalsize \itshape 141980 Dubna, Moscow region, Russia} \\ [5pt]
{\normalsize \itshape D\'epartement de Math\'ematiques, Universit\'e d'Angers, 49045 Angers, France}}
\title{Transition function for the Toda chain model}
\date{}

\begin{document}

\maketitle

\begin{abstract}
The method of $\Lambda$-operators developed by S.\,Derkachov, G.\,Korchemsky, A.\,Ma\-na\-shov is
applied to a derivation of eigenfunctions for the open Toda chain.
The Sklyanin measure is reproduced using diagram technique developed
for these $\Lambda$-operators. The properties of the
$\Lambda$-operators are studied. This approach to the open Toda
chain eigenfunctions reproduces Gauss-Givental representation for
these eigenfunctions.
%
%
\end{abstract}


\section{Introduction}

This work was inspired by the article~\cite{Derkachov} devoted to
the Separation of Variables (SoV) method for $XXX$-model. The main
idea of this method is to find an integral transformation such
that eigenfunctions of quantum integrals of motion in new
variables becomes the product of functions of one variable~\cite{Sklyanin}. If
everyone of these functions satisfies the Baxter equation, then
the initial multivariable function becomes an eigenfunction. The
kernel of this transform is called a transition function and can
be constructed as consecutive application of operators
$\Lambda_k(u)$:
$\Lambda_N(\gamma_1)\Lambda_{N-1}(\gamma_2)\ldots\Lambda_1(\gamma_N)$.
Every operator $\Lambda_k(u)$ is an integral
transformation, which maps a function of $k-1$ variables onto
function of $k$ variables. The properties of the transition
function can be translated to the properties of these operators (see
section~\ref{Pr_Lambda}). \\

The transition function for the $N$-particle periodic Toda chain
was obtained in the works~\cite{Gutzwiller}, \cite{Sklyanin}, \cite{Pasquier}
\cite{Kharchev_P}. The transition function in this case is
proportional to the eigenfunction of the $(N-1)$-particle open Toda chain, with a
factor depending on the $N$-th particle coordinate. We apply
methods of the paper~\cite{Derkachov} to obtain
these eigenfunctions as a product of $\Lambda$-operators reproducing the Gauss-Givental formula~\cite{Kharchev_GG}, \cite{Givental}. \\

This form of eigenfunctions of the open Toda chain leads to an
integral representation that appeared first in~\cite{Givental}
employing a different approach. Recently it was interpreted from a
group-theoretical point of view using the Gauss decomposition of
$GL(N,\mathbb R)$~\cite{Kharchev_GG}. Therefore, this integral
representation of the eigenfunctions for the open Toda chain is
called a Gauss-Givental representation. \\

The method of a triangulation of the Lax matrix described
in~\cite{Pasquier} was used in~\cite{Derkachov}. We also use a
triangulation, which is implemented by a gauge transformation
parameterized by variables $y_0,\ldots,y_N$. In the periodic case
one has to impose the condition $y_0=y_N$ and the method described
in~\cite{Pasquier} produces Baxter's $Q$-operators~\cite{Baxter} for the
periodic Toda chain model. Following~\cite{Derkachov} we impose a
different boundary condition: $y_0\to -\infty$, $y_N\to +\infty$
to construct $\Lambda$-operator. Thus $\Lambda$-operator and
Baxter's Q-operator for the periodic Toda chain correspond to the
different choice of the boundary conditions in the method of
triangulation of the Lax matrices. \\

To describe the construction of eigenfunctions for open Toda
chain we develop a kind of the Feynman diagram technique similar to
one exploited in~\cite{Derkachov}. It allows to reduce
calculations with kernels of $\Lambda$-operators to simple
manipulations with diagrams. \\

The article is organized as follows. In section~\ref{OTCh} we
recall a definition of the open Toda chain model following~\cite{Kharchev_O, Kharchev_OP}.
Section~\ref{Eig_OTCh} is devoted to a description of
eigenfunctions in terms of the product of $\Lambda$-operators
and formulation of a diagram technique developed in~\cite{Derkachov}. In
section~\ref{Int_meas} we use this technique in order to prove
that eigenfunctions satisfy an orthogonality condition. As a by-product
of this calculation we obtain a Sklyanin measure, which is necessary
to prove a completeness condition.
Section~\ref{Pr_Lambda} is devoted to algebraic properties of
$\Lambda$-operators.

\section{Open Toda chain model}
\label{OTCh}

 The quantum $N$-particle open Toda chain is a one-dimensional model with
the exponential interaction between the nearest particles.
The hamiltonian of the system is equal to
\begin{equation}
 H=\frac12\sum_{n=1}^N p_n^2+\sum_{n=1}^{N-1} e^{x_n-x_{n+1}},
\end{equation}
where $p_n\hm=-i\hbar\dfrac{\partial}{\partial x_n}$ is an operator of momentum
for the $n$-th particle. Due to the translational invariance, the total momentum
\begin{equation}
 P=\sum_{n=1}^N p_n
\end{equation}
commutes with the hamiltonian, i.e. it is also an integral of motion.
There are $N$ functionally independent integrals of motion for
this system. It is relevant to use the $R$-matrix formalism to find them.
First of all, introduce the Lax operator of the Toda chain
\begin{equation} \label{Ln}
 L_n(u)=
  \begin{pmatrix}
   u-p_n & e^{-x_n} \\
   -e^{x_n} & 0
  \end{pmatrix}, \qquad n=1,\ldots,N,
\end{equation}
and monodromy matrix for the $N$-particle Toda chain
\begin{equation} \label{TN}
 T_N(u)=L_N(u)\cdots L_1(u)=
  \begin{pmatrix}
   A_N(u) & B_N(u) \\
   C_N(u) & D_N(u)
  \end{pmatrix},
\end{equation}
where $u$ is a spectral parameter. \\

The following recurrent relations, which are direct consequence of this definition,
will be useful below:
\begin{gather}
 A_N(u)=(u-p_N)A_{N-1}(u)+e^{-x_N}C_{N-1}(u), \label{Arec} \\
 C_N(u)=-e^{x_N}A_{N-1}(u), \label{Crec} \\
 A_N(u)=(u-p_N)A_{N-1}(u)-e^{x_{N-1}-x_N}A_{N-2}(u).  \label{AArec}
\end{gather}
These relations show that $A_N(u)$ and $C_N(u)$ are polynomials in
$u$ of degree $N$ and $N-1$ respectively. Analogously, $B_N(u)$
and $D_N(u)$ have degree $N-1$ and $N-2$. \\

The monodromy matrix satisfies to the quantum $RTT$-relation
\begin{equation} \label{RTT}
  R(u-v)\,(T_N(u)\otimes I)(I\otimes T_N(v))=
  (I\otimes T_N(v))(T_N(u)\otimes I)\,R(u-v)
\end{equation}
with the rational $R$-matrix
\begin{equation} \label{RP}
 R(u)=I\otimes I+\frac{i\hbar}{u}\,\mathcal P,
\end{equation}
where $\mathcal P$ is a permutation matrix: $\mathcal P_{ij,kl}\hm=\delta_{il}\delta_{jk}$.

Rewriting~\eqref{RTT} by entries one obtains, in particular, the
relation
\begin{equation} \label{RTTAA}
 A_N(u)A_N(v)=A_N(v)A_N(u).
\end{equation}

This means that $A_N(u)$ is a generation function of the integrals
of motion of the integrable system with $N$ degree of freedom.
Explicit calculations show that these are integrals for open Toda
chain model:
\begin{gather}
 A_N(u)=\sum_{k=0}^N (-1)^k u^{N-k} H_k, \\
 H_0=1, \quad H_1=P, \quad H_2=\frac12 P^2-H, \\
 [H_k,H_j]=0. \label{commHH}
\end{gather}

By virtue of~\eqref{commHH} there exist common eigenfunctions of the integrals $H_k$
corresponding to the eigenvalues $E_k$. They are defining by the equation
\begin{equation} \label{psi_def_E}
 A_N(u)\psi_E(x)= a_N(u;E)\psi_E(x),
\end{equation}
where
\begin{equation*}
 a_N(u;E)=\sum_{k=0}^N (-1)^k u^{N-k} E_k,
\end{equation*}
$E_0=1$, $E=(E_1,\ldots,E_N)$, $x=(x_1,\ldots,x_N)$. Representing
eigenvalues $E_k$ as symmetric combinations
\begin{equation} \label{Ek}
 E_k=\sum\limits_{j_1<\ldots<j_k}\gamma_{j_1}\ldots\gamma_{j_k}
\end{equation}
of real variables $\gamma=(\gamma_1,\ldots,\gamma_N)$, one can rewrite equation~\eqref{psi_def_E} as follows
\begin{equation} \label{psi_def}
 A_N(u)\psi_{\gamma}(x)= \prod_{j=1}^N(u-\gamma_j)\psi_{\gamma}(x).
\end{equation}

We also require the solutions of the equation~\eqref{psi_def} to be rapidly decaying in region $x_i\gg x_{i+1}$. This asymptotic properties of solutions correspond to the condition that the $i$-th particle is situated on the left of the $(i+1)$-th particle.

\section{Eigenfunctions of the open Toda chain}
\label{Eig_OTCh}

In this section we shall find eigenfunctions of the open Toda
chain defined in the previous section by the
equation~\eqref{psi_def}. This equation is equivalent to the
system of $N$ equations
\begin{equation} \label{Apsi0}
 A_N(\gamma_j)\psi_{\gamma}(x)=0, \qquad j=1,\ldots,N.
\end{equation}
The eigenvalues~\eqref{Ek} are invariant under the permutations of $\gamma_1,\ldots,\gamma_N$.
Therefore, it is reasonably to require the invariance of eigenfunction under these permutations, which we shall call the Weyl invariance:
\begin{equation} \label{Winv}
 \psi_{\sigma(\gamma)}(x)=\psi_{\gamma}(x), \qquad \text{for all } \sigma\in S_N,
\end{equation}
where $S_N$ is a permutation group and $\sigma(\gamma)
=(\gamma_{\sigma(1)},\ldots, \gamma_{\sigma(N)})$. \\

It is sufficiently to find a Weyl invariant solution of the unique
equation
\begin{equation} \label{Agamma1psi0}
 A_N(\gamma_1)\psi_{\gamma}(x)=0,
\end{equation}
which will be a solution for the whole system~\eqref{Apsi0} due to
its Weyl invariance. \\

To solve the last equation we shall consider a gauge transformation of the Lax operators
\begin{equation} \label{MLM}
 \widetilde L_n(u)=M_{n}L_n(u)M_{n-1}^{-1}, \qquad n=1,\ldots,N
\end{equation}
by the matrices
\begin{equation}
 M_n=\begin{pmatrix}
   1 & 0 \\
   ie^{y_n} & 1
  \end{pmatrix}, \qquad n=0,\ldots,N.
\end{equation}
The deformed $N$-particle monodromy matrix is
\begin{equation}
 \widetilde T_N(u)\equiv
  \begin{pmatrix}
   \widetilde A_N(u) & \widetilde B_N(u) \\
   \widetilde C_N(u) & \widetilde D_N(u)
  \end{pmatrix}
  =\widetilde L_N(u)\cdots\widetilde L_1(u)=M_N T_N(u) M_0^{-1}.
\end{equation}
In particular, we have
\begin{gather}
 \widetilde L_n(u)_{21}=ie^{y_n}(u-p_n-ie^{y_{n-1}-x_n}+ie^{x_n-y_n}), \label{L21_til}\\
 \widetilde C_N(u)=ie^{y_N} A_N(u)+e^{y_N+y_0}B_N(u)+C_N(u)-ie^{y_0}D_N(u). \label{C_til}
\end{gather}
Here $\widetilde L_n(u)_{21}$ is a lower off-diagonal entry of the matrix $\widetilde L_n(u)$. \\

Let us consider the auxiliary equation
\begin{equation}
 \widetilde L_n(u)_{21}w_n(u)=0,
\end{equation}
which has the following solution
\begin{equation}
 w_n(u)=\exp\Bigl\{\frac{i}{\hbar}u(x_n-y_{n-1})-\frac{1}{\hbar}e^{y_{n-1}-x_n}-\frac{1}{\hbar}e^{x_n-y_n}\Bigr\}.
\end{equation}
It is clear that the function
\begin{equation}
 W_u(x;y)\hm=\prod\limits_{n=1}^N w_n(u)
 =\exp\sum\limits_{n=1}^N\Bigl\{\frac{i}{\hbar}u(x_n-y_{n-1})-\frac{1}{\hbar}e^{y_{n-1}-x_n}-\frac{1}{\hbar}e^{x_n-y_n}\Bigr\}
\end{equation}
is a solution to the equation
\begin{equation} \label{C_tilW}
 \widetilde C_N(u) W_u(x;y)=0.
\end{equation}
In the limit $y_0\to -\infty$, $y_N\to +\infty$ the formula~\eqref{C_til} gives us
the equality
\begin{equation} \label{ANlim}
A_N(u)\hm=-i\lim\limits_{\substack{y_0\to -\infty \\ y_N\to +\infty}} e^{-y_N}\widetilde C_N(u).
\end{equation}
Therefore, multiplying the equation~\eqref{C_tilW} by
$-ie^{-y_N}e^{\frac{i}{\hbar}u(y_0+y_N)}$, taking the same limit
as in~\eqref{ANlim} and setting $u\hm=\gamma_1$ we arrive to the
equation~\eqref{Agamma1psi0} with the solution
$\psi_{\gamma}(x)\hm=\Lambda_{\gamma_1}(x;y)$, where
\begin{equation} \label{Lamb}
 \begin{split}
 \Lambda_u(x;y)=\Lambda_u(x_1,\ldots,x_N;y_1,\ldots,y_{N-1})
     &=\lim\limits_{\substack{y_0\to -\infty \\ y_N\to +\infty}}
        e^{\frac{i}{\hbar}u(y_0+y_N)}W_{u}(x;y)=\\
     =\exp\Bigl\{\frac{i}{\hbar}u(\sum\limits_{n=1}^N x_n-&\sum\limits_{n=1}^{N-1}y_n)
        -\frac{1}{\hbar}\sum\limits_{n=1}^{N-1}(e^{y_n-x_{n+1}}+e^{x_n-y_n})\Bigr\}.
 \end{split}
\end{equation}

Let $\Lambda_N(u)$ be an operator with the kernel
$\Lambda_u(x_1,\ldots,x_N;y_1,\ldots,y_{N-1})$, i.e.
\begin{equation} \label{Lambf}
 (\Lambda_N(u)\cdot f)(x)
   =\int\limits_{\mathbb R^{N-1}}dy\,\Lambda_u(x_1,\ldots,x_N;y_1,\ldots,y_{N-1})f(y).
\end{equation}
Since $|\Lambda_u(x;y)|$ decays at $y_i\to\pm\infty$ the integral in the right hand side of~\eqref{Lambf} converges absolutely for the functions $f(y)$ increasing sufficiently slowly in the infinity.
Setting $u=\gamma_1$ in~\eqref{Lambf} we obtain a solution to~\eqref{Agamma1psi0}.
Since $|\Lambda_u(x;y)|$ decays in the region $x_i\gg x_{i+1}$ this solution has the requiring asymptotic properties.

\begin{theorem} \label{Th_psi}
 The following solution to the equation~\eqref{Agamma1psi0}
\begin{equation} \label{psi}
 \psi_{\gamma}(x)=(\Lambda_N(\gamma_1)\cdots\Lambda_2(\gamma_{N-1})\Lambda_1(\gamma_N)\cdot 1)(x_1,\ldots,x_N),
\end{equation}
where $(\Lambda_1(\gamma_N)\cdot
1)(x_1)=e^{\frac{i}{\hbar}\gamma_N x_1}$, is Weyl invariant, i.e.
satisfies to the condition~\eqref{Winv}, and, therefore, is a
solution to the equation~\eqref{psi_def} rapidly decaying in the region $x_i\gg x_{i+1}$.
\end{theorem}

\noindent{\bfseries Proof.} It is sufficient to establish the invariance under the elementary
permutations, i.e. we need to check the equality
\begin{equation} \label{LgammaLgamma}
 \begin{split}
  &\int\limits_{\mathbb R^{N-n+1}}dy\,\Lambda_{\gamma_{n-1}}(x_1,\ldots,x_{N-n+2};y_1,\ldots,y_{N-n+1})
                            \Lambda_{\gamma_n}(y_1,\ldots,y_{N-n+1};z_1,\ldots,z_{N-n})= \\
  &=\int\limits_{\mathbb R^{N-n+1}}dy\,\Lambda_{\gamma_n}(x_1,\ldots,x_{N-n+2};y_1,\ldots,y_{N-n+1})
                            \Lambda_{\gamma_{n-1}}(y_1,\ldots,y_{N-n+1};z_1,\ldots,z_{N-n}),
 \end{split}
\end{equation}
for $n=1,\ldots,N-1$. \\

\hspace{\fill} \setlength{\unitlength}{0.00087489in}
\begingroup\makeatletter\ifx\SetFigFont\undefined%
\gdef\SetFigFont#1#2#3#4#5{%
  \reset@font\fontsize{#1}{#2pt}%
  \fontfamily{#3}\fontseries{#4}\fontshape{#5}%
  \selectfont}%
\fi\endgroup%
{\renewcommand{\dashlinestretch}{30}
\begin{picture}(781,1336)(0,-10)
\path(60,343)(330,883)
\path(303.167,762.252)(330.000,883.000)(249.502,789.085)
\path(60,343)(510,1243)
\put(150,298){\makebox(0,0)[lb]{\smash{{\SetFigFont{12}{14.4}{\rmdefault}{\mddefault}{\updefault}$x_k$}}}}
\put(600,1198){\makebox(0,0)[lb]{\smash{{\SetFigFont{12}{14.4}{\rmdefault}{\mddefault}{\updefault}$y_k$}}}}
\put(5,-23){\makebox(0,0)[lb]{\smash{{\SetFigFont{14}{14.4}{\rmdefault}{\mddefault}{\updefault}fig.
1a}}}}
\end{picture}
} \hspace{\fill} \setlength{\unitlength}{0.00087489in}
\begingroup\makeatletter\ifx\SetFigFont\undefined%
\gdef\SetFigFont#1#2#3#4#5{%
  \reset@font\fontsize{#1}{#2pt}%
  \fontfamily{#3}\fontseries{#4}\fontshape{#5}%
  \selectfont}%
\fi\endgroup%
{\renewcommand{\dashlinestretch}{30}
\begin{picture}(736,1318)(0,-10)
\path(465,343)(15,1243) \path(465,343)(15,1243)
\path(465,343)(195,883) \path(465,343)(195,883)
\whiten\path(275.498,789.085)(195.000,883.000)(221.833,762.252)(264.765,743.469)(275.498,789.085)
\put(105,1198){\makebox(0,0)[lb]{\smash{{\SetFigFont{12}{14.4}{\rmdefault}{\mddefault}{\updefault}$y_k$}}}}
\put(555,298){\makebox(0,0)[lb]{\smash{{\SetFigFont{12}{14.4}{\rmdefault}{\mddefault}{\updefault}$x_{k+1}$}}}}
\put(15,793){\makebox(0,0)[lb]{\smash{{\SetFigFont{10}{14.4}{\rmdefault}{\mddefault}{\updefault}$u$}}}}
\put(50,-23){\makebox(0,0)[lb]{\smash{{\SetFigFont{14}{14.4}{\rmdefault}{\mddefault}{\updefault}fig.
1b}}}}
\end{picture}
} \hspace{\fill} \setlength{\unitlength}{0.00087489in}
\begingroup\makeatletter\ifx\SetFigFont\undefined%
\gdef\SetFigFont#1#2#3#4#5{%
  \reset@font\fontsize{#1}{#2pt}%
  \fontfamily{#3}\fontseries{#4}\fontshape{#5}%
  \selectfont}%
\fi\endgroup%
{\renewcommand{\dashlinestretch}{30}
\begin{picture}(736,913)(0,-10)
\put(555,298){\makebox(0,0)[lb]{\smash{{\SetFigFont{12}{14.4}{\rmdefault}{\mddefault}{\updefault}$x_k$}}}}
\path(465,343)(195,883) \path(465,343)(195,883)
\whiten\path(275.498,789.085)(195.000,883.000)(221.833,762.252)(275.498,789.085)
\put(15,793){\makebox(0,0)[lb]{\smash{{\SetFigFont{10}{14.4}{\rmdefault}{\mddefault}{\updefault}$u$}}}}
\put(160,-23){\makebox(0,0)[lb]{\smash{{\SetFigFont{14}{14.4}{\rmdefault}{\mddefault}{\updefault}fig.
1c}}}}
\end{picture}
} \hspace{\fill} \setlength{\unitlength}{0.00087489in}
\begingroup\makeatletter\ifx\SetFigFont\undefined%
\gdef\SetFigFont#1#2#3#4#5{%
  \reset@font\fontsize{#1}{#2pt}%
  \fontfamily{#3}\fontseries{#4}\fontshape{#5}%
  \selectfont}%
\fi\endgroup%
{\renewcommand{\dashlinestretch}{30}
\begin{picture}(522,913)(0,-10)
\path(105,838)(330,388) \path(105,838)(330,388)
\path(330,388)(285,478) \path(330,388)(285,478)
\whiten\path(365.498,384.085)(285.000,478.000)(311.833,357.252)(365.498,384.085)
\put(195,793){\makebox(0,0)[lb]{\smash{{\SetFigFont{12}{14.4}{\rmdefault}{\mddefault}{\updefault}$x_k$}}}}
\put(105,388){\makebox(0,0)[lb]{\smash{{\SetFigFont{10}{14.4}{\rmdefault}{\mddefault}{\updefault}$u$}}}}
\put(05,-23){\makebox(0,0)[lb]{\smash{{\SetFigFont{14}{14.4}{\rmdefault}{\mddefault}{\updefault}fig.
1d}}}}
\end{picture}
} \hspace{\fill} \setlength{\unitlength}{0.00087489in}
\begingroup\makeatletter\ifx\SetFigFont\undefined%
\gdef\SetFigFont#1#2#3#4#5{%
  \reset@font\fontsize{#1}{#2pt}%
  \fontfamily{#3}\fontseries{#4}\fontshape{#5}%
  \selectfont}%
\fi\endgroup%
{\renewcommand{\dashlinestretch}{30}
\begin{picture}(589,1093)(0,-10)
\path(195,343)(195,1018) \path(195,343)(195,793)
\whiten\path(225.000,673.000)(195.000,793.000)(165.000,673.000)(195.000,709.000)(225.000,673.000)
\put(285,298){\makebox(0,0)[lb]{\smash{{\SetFigFont{12}{14.4}{\rmdefault}{\mddefault}{\updefault}$x_{k+1}$}}}}
\put(285,973){\makebox(0,0)[lb]{\smash{{\SetFigFont{12}{14.4}{\rmdefault}{\mddefault}{\updefault}$z_k$}}}}
\put(05,-23){\makebox(0,0)[lb]{\smash{{\SetFigFont{14}{14.4}{\rmdefault}{\mddefault}{\updefault}fig.
1e}}}}
\put(15,658){\makebox(0,0)[lb]{\smash{{\SetFigFont{10}{12.0}{\rmdefault}{\mddefault}{\updefault}$u$}}}}
\end{picture}
} \hspace{\fill} \\

To do it we shall use a diagram technique introduced
in~\cite{Derkachov}. Let us denote the function
$I(x_k,y_k)\hm=\exp\Bigl\{-\dfrac{1}{\hbar}e^{x_k-y_k}\Bigr\}$ by
a line pictured in the fig.~1a, the function
$J_u(x_{k+1},y_k)\hm=\exp\Bigl\{\dfrac{i}{\hbar}u(x_{k+1}-y_k)\hm-\dfrac{1}{\hbar}e^{y_k-x_{k+1}}\Bigr\}$
by a fig.~1b, the function
$Z_u(x_k)\hm=\exp\Bigl\{\dfrac{i}{\hbar}u x_k\Bigr\}$ by a
fig.~1c, the function
$Z_u^{-1}(x_k)\hm=\exp\Bigl\{-\dfrac{i}{\hbar}u x_k\Bigr\}$ by a
fig.~1d and the function
$Y_u(x_{k+1},z_k)\hm=\Bigl(1+e^{x_{k+1}-z_k}\Bigr)^{\tfrac{i}{\hbar}u}$
by a fig.~1e. \\

\hspace{\fill} \setlength{\unitlength}{0.00087489in}
\begingroup\makeatletter\ifx\SetFigFont\undefined%
\gdef\SetFigFont#1#2#3#4#5{%
  \reset@font\fontsize{#1}{#2pt}%
  \fontfamily{#3}\fontseries{#4}\fontshape{#5}%
  \selectfont}%
\fi\endgroup%
{\renewcommand{\dashlinestretch}{30}
\begin{picture}(4291,4293)(0,-10)
\path(465,105)(735,645)
\path(708.167,524.252)(735.000,645.000)(654.502,551.085)
\path(465,105)(915,1005) \path(1365,105)(915,1005)
\path(1365,105)(915,1005) \path(1365,105)(1095,645)
\path(1365,105)(1095,645)
\whiten\path(1175.498,551.085)(1095.000,645.000)(1121.833,524.252)(1164.765,505.469)(1175.498,551.085)
\path(2265,105)(2535,645)
\path(2508.167,524.252)(2535.000,645.000)(2454.502,551.085)
\path(2265,105)(2715,1005) \path(3165,105)(2715,1005)
\path(3165,105)(2715,1005) \path(3165,105)(2895,645)
\path(3165,105)(2895,645)
\whiten\path(2975.498,551.085)(2895.000,645.000)(2921.833,524.252)(2964.765,505.469)(2975.498,551.085)
\path(2715,1005)(2265,1905) \path(2715,1005)(2265,1905)
\path(2715,1005)(2445,1545) \path(2715,1005)(2445,1545)
\whiten\path(2525.498,1451.085)(2445.000,1545.000)(2471.833,1424.252)(2514.765,1405.469)(2525.498,1451.085)
\put(915,555){\makebox(0,0)[lb]{\smash{{\SetFigFont{10}{14.4}{\rmdefault}{\mddefault}{\updefault}$\gamma_1$}}}}
\put(2715,555){\makebox(0,0)[lb]{\smash{{\SetFigFont{10}{14.4}{\rmdefault}{\mddefault}{\updefault}$\gamma_1$}}}}
\put(3615,555){\makebox(0,0)[lb]{\smash{{\SetFigFont{10}{14.4}{\rmdefault}{\mddefault}{\updefault}$\gamma_1$}}}}
\put(3615,555){\makebox(0,0)[lb]{\smash{{\SetFigFont{10}{14.4}{\rmdefault}{\mddefault}{\updefault}$\gamma_1$}}}}
\put(3165,1455){\makebox(0,0)[lb]{\smash{{\SetFigFont{10}{14.4}{\rmdefault}{\mddefault}{\updefault}$\gamma_2$}}}}
\put(1815,2355){\makebox(0,0)[lb]{\smash{{\SetFigFont{10}{14.4}{\rmdefault}{\mddefault}{\updefault}$\gamma_3$}}}}
\put(2715,2355){\makebox(0,0)[lb]{\smash{{\SetFigFont{10}{14.4}{\rmdefault}{\mddefault}{\updefault}$\gamma_3$}}}}
\path(915,1005)(1185,1545)
\path(1158.167,1424.252)(1185.000,1545.000)(1104.502,1451.085)
\path(915,1005)(1365,1905) \path(1365,1905)(1635,2445)
\path(1608.167,2324.252)(1635.000,2445.000)(1554.502,2351.085)
\path(1365,1905)(1815,2805) \path(2265,1905)(1815,2805)
\path(2265,1905)(1815,2805) \path(2265,1905)(1995,2445)
\path(2265,1905)(1995,2445)
\whiten\path(2075.498,2351.085)(1995.000,2445.000)(2021.833,2324.252)(2064.765,2305.469)(2075.498,2351.085)
\path(1815,2805)(2085,3345)
\path(2058.167,3224.252)(2085.000,3345.000)(2004.502,3251.085)
\path(1815,2805)(2265,3705) \path(2715,2805)(2265,3705)
\path(2715,2805)(2265,3705) \path(2715,2805)(2445,3345)
\path(2715,2805)(2445,3345)
\whiten\path(2525.498,3251.085)(2445.000,3345.000)(2471.833,3224.252)(2514.765,3205.469)(2525.498,3251.085)
\path(2265,1905)(2535,2445)
\path(2508.167,2324.252)(2535.000,2445.000)(2454.502,2351.085)
\path(2265,1905)(2715,2805) \path(3165,1905)(2715,2805)
\path(3165,1905)(2715,2805) \path(3165,1905)(2895,2445)
\path(3165,1905)(2895,2445)
\whiten\path(2975.498,2351.085)(2895.000,2445.000)(2921.833,2324.252)(2964.765,2305.469)(2975.498,2351.085)
\path(2715,1005)(2985,1545)
\path(2958.167,1424.252)(2985.000,1545.000)(2904.502,1451.085)
\path(2715,1005)(3165,1905) \path(3615,1005)(3165,1905)
\path(3615,1005)(3165,1905) \path(3615,1005)(3345,1545)
\path(3615,1005)(3345,1545)
\whiten\path(3425.498,1451.085)(3345.000,1545.000)(3371.833,1424.252)(3414.765,1405.469)(3425.498,1451.085)
\path(3165,105)(3435,645)
\path(3408.167,524.252)(3435.000,645.000)(3354.502,551.085)
\path(3165,105)(3615,1005) \path(4065,105)(3615,1005)
\path(4065,105)(3615,1005) \path(4065,105)(3795,645)
\path(4065,105)(3795,645)
\whiten\path(3875.498,551.085)(3795.000,645.000)(3821.833,524.252)(3864.765,505.469)(3875.498,551.085)
\put(15,555){\makebox(0,0)[lb]{\smash{{\SetFigFont{10}{14.4}{\rmdefault}{\mddefault}{\updefault}$\gamma_1$}}}}
\put(555,60){\makebox(0,0)[lb]{\smash{{\SetFigFont{12}{14.4}{\rmdefault}{\mddefault}{\updefault}$x_1$}}}}
\path(465,105)(195,645) \path(465,105)(195,645)
\whiten\path(275.498,551.085)(195.000,645.000)(221.833,524.252)(275.498,551.085)
\path(1365,105)(1635,645)
\path(1608.167,524.252)(1635.000,645.000)(1554.502,551.085)
\path(1365,105)(1815,1005) \path(2265,105)(1815,1005)
\path(2265,105)(1815,1005) \path(2265,105)(1995,645)
\path(2265,105)(1995,645)
\whiten\path(2075.498,551.085)(1995.000,645.000)(2021.833,524.252)(2064.765,505.469)(2075.498,551.085)
\put(1815,1005){\blacken\ellipse{46}{46}}
\put(1815,1005){\ellipse{46}{46}} \path(1815,1005)(2085,1545)
\path(2058.167,1424.252)(2085.000,1545.000)(2004.502,1451.085)
\path(1815,1005)(2265,1905)
\put(1815,555){\makebox(0,0)[lb]{\smash{{\SetFigFont{10}{14.4}{\rmdefault}{\mddefault}{\updefault}$\gamma_1$}}}}
\put(1365,1455){\makebox(0,0)[lb]{\smash{{\SetFigFont{10}{14.4}{\rmdefault}{\mddefault}{\updefault}$\gamma_2$}}}}
\put(2265,1455){\makebox(0,0)[lb]{\smash{{\SetFigFont{10}{14.4}{\rmdefault}{\mddefault}{\updefault}$\gamma_2$}}}}
\path(1815,1005)(1365,1905) \path(1815,1005)(1365,1905)
\path(1815,1005)(1545,1545) \path(1815,1005)(1545,1545)
\whiten\path(1625.498,1451.085)(1545.000,1545.000)(1571.833,1424.252)(1614.765,1405.469)(1625.498,1451.085)
\put(915,1005){\blacken\ellipse{46}{46}}
\put(915,1005){\ellipse{46}{46}}
\put(2715,1005){\blacken\ellipse{46}{46}}
\put(2715,1005){\ellipse{46}{46}}
\put(1815,2805){\blacken\ellipse{46}{46}}
\put(1815,2805){\ellipse{46}{46}}
\put(2265,3705){\blacken\ellipse{46}{46}}
\put(2265,3705){\ellipse{46}{46}}
\put(2715,2805){\blacken\ellipse{46}{46}}
\put(2715,2805){\ellipse{46}{46}}
\put(3165,1905){\blacken\ellipse{46}{46}}
\put(3165,1905){\ellipse{46}{46}}
\put(3615,1005){\blacken\ellipse{46}{46}}
\put(3615,1005){\ellipse{46}{46}}
\put(2265,1905){\blacken\ellipse{46}{46}}
\put(2265,1905){\ellipse{46}{46}}
\put(1365,1905){\blacken\ellipse{46}{46}}
\put(1365,1905){\ellipse{46}{46}} \path(915,1005)(645,1545)
\path(915,1005)(645,1545)
\whiten\path(725.498,1451.085)(645.000,1545.000)(671.833,1424.252)(725.498,1451.085)
\path(1365,1905)(1095,2445) \path(1365,1905)(1095,2445)
\whiten\path(1175.498,2351.085)(1095.000,2445.000)(1121.833,2324.252)(1175.498,2351.085)
\path(1815,2805)(1545,3345) \path(1815,2805)(1545,3345)
\whiten\path(1625.498,3251.085)(1545.000,3345.000)(1571.833,3224.252)(1625.498,3251.085)
\path(2265,3705)(1995,4245) \path(2265,3705)(1995,4245)
\whiten\path(2075.498,4151.085)(1995.000,4245.000)(2021.833,4124.252)(2075.498,4151.085)
\put(4110,60){\makebox(0,0)[lb]{\smash{{\SetFigFont{12}{14.4}{\rmdefault}{\mddefault}{\updefault}$x_5$}}}}
\put(1905,2760){\makebox(0,0)[lb]{\smash{{\SetFigFont{12}{14.4}{\rmdefault}{\mddefault}{\updefault}$f_1$}}}}
\put(2805,2760){\makebox(0,0)[lb]{\smash{{\SetFigFont{12}{14.4}{\rmdefault}{\mddefault}{\updefault}$f_2$}}}}
\put(3255,60){\makebox(0,0)[lb]{\smash{{\SetFigFont{12}{14.4}{\rmdefault}{\mddefault}{\updefault}$x_4$}}}}
\put(1005,960){\makebox(0,0)[lb]{\smash{{\SetFigFont{12}{14.4}{\rmdefault}{\mddefault}{\updefault}$y_1$}}}}
\put(2805,960){\makebox(0,0)[lb]{\smash{{\SetFigFont{12}{14.4}{\rmdefault}{\mddefault}{\updefault}$y_3$}}}}
\put(3705,960){\makebox(0,0)[lb]{\smash{{\SetFigFont{12}{14.4}{\rmdefault}{\mddefault}{\updefault}$y_4$}}}}
\put(3255,1860){\makebox(0,0)[lb]{\smash{{\SetFigFont{12}{14.4}{\rmdefault}{\mddefault}{\updefault}$z_3$}}}}
\put(2355,3660){\makebox(0,0)[lb]{\smash{{\SetFigFont{12}{14.4}{\rmdefault}{\mddefault}{\updefault}$g_1$}}}}
\put(2265,3255){\makebox(0,0)[lb]{\smash{{\SetFigFont{10}{14.4}{\rmdefault}{\mddefault}{\updefault}$\gamma_4$}}}}
\put(465,1455){\makebox(0,0)[lb]{\smash{{\SetFigFont{10}{14.4}{\rmdefault}{\mddefault}{\updefault}$\gamma_2$}}}}
\put(915,2355){\makebox(0,0)[lb]{\smash{{\SetFigFont{10}{14.4}{\rmdefault}{\mddefault}{\updefault}$\gamma_3$}}}}
\put(1365,3255){\makebox(0,0)[lb]{\smash{{\SetFigFont{10}{14.4}{\rmdefault}{\mddefault}{\updefault}$\gamma_4$}}}}
\put(1815,4155){\makebox(0,0)[lb]{\smash{{\SetFigFont{10}{14.4}{\rmdefault}{\mddefault}{\updefault}$\gamma_5$}}}}
\put(2355,1860){\makebox(0,0)[lb]{\smash{{\SetFigFont{12}{14.4}{\rmdefault}{\mddefault}{\updefault}$z_2$}}}}
\put(1455,60){\makebox(0,0)[lb]{\smash{{\SetFigFont{12}{14.4}{\rmdefault}{\mddefault}{\updefault}$x_2$}}}}
\put(2355,60){\makebox(0,0)[lb]{\smash{{\SetFigFont{12}{14.4}{\rmdefault}{\mddefault}{\updefault}$x_3$}}}}
\put(1905,960){\makebox(0,0)[lb]{\smash{{\SetFigFont{12}{14.4}{\rmdefault}{\mddefault}{\updefault}$y_2$}}}}
\put(1455,1860){\makebox(0,0)[lb]{\smash{{\SetFigFont{12}{14.4}{\rmdefault}{\mddefault}{\updefault}$z_1$}}}}
\end{picture}
}
 \hspace{\fill} \\
\vspace{-2pt} \hspace{77mm} {\large fig. 2}

The function~\eqref{psi} can be represented in these graphical
notations. For the case $N=5$ it is pictured in fig.~2, where bold
bullets $\bullet$ signify that we integrate over corresponding
variables.

\begin{lemma}
 The equalities represented in the figures~3a, 3b, 3c are valid.
\end{lemma}
\vspace{7pt}

\hspace{\fill} \setlength{\unitlength}{0.00087489in}
\begingroup\makeatletter\ifx\SetFigFont\undefined%
\gdef\SetFigFont#1#2#3#4#5{%
  \reset@font\fontsize{#1}{#2pt}%
  \fontfamily{#3}\fontseries{#4}\fontshape{#5}%
  \selectfont}%
\fi\endgroup%
{\renewcommand{\dashlinestretch}{30}
\begin{picture}(4186,1980)(0,-10)
\put(555,1005){\makebox(0,0)[rb]{\smash{{\SetFigFont{10}{12.0}{\rmdefault}{\mddefault}{\updefault}\makebox[0.8\width][s]{$(\gamma_{n-1}-\gamma_n)$}}}}}
\path(690,105)(960,645)
\path(933.167,524.252)(960.000,645.000)(879.502,551.085)
\path(690,105)(1140,1005) \path(1590,105)(1140,1005)
\path(1590,105)(1140,1005) \path(1590,105)(1320,645)
\path(1590,105)(1320,645)
\whiten\path(1400.498,551.085)(1320.000,645.000)(1346.833,524.252)(1389.765,505.469)(1400.498,551.085)
\put(1140,1005){\blacken\ellipse{46}{46}}
\put(1140,1005){\ellipse{46}{46}}
\put(780,1860){\makebox(0,0)[lb]{\smash{{\SetFigFont{12}{14.4}{\rmdefault}{\mddefault}{\updefault}$z_{k-1}$}}}}
\put(1680,1860){\makebox(0,0)[lb]{\smash{{\SetFigFont{12}{14.4}{\rmdefault}{\mddefault}{\updefault}$z_k$}}}}
\put(1230,960){\makebox(0,0)[lb]{\smash{{\SetFigFont{12}{14.4}{\rmdefault}{\mddefault}{\updefault}$y_k$}}}}
\put(780,60){\makebox(0,0)[lb]{\smash{{\SetFigFont{12}{14.4}{\rmdefault}{\mddefault}{\updefault}$x_k$}}}}
\put(1680,60){\makebox(0,0)[lb]{\smash{{\SetFigFont{12}{14.4}{\rmdefault}{\mddefault}{\updefault}$x_{k+1}$}}}}
\path(1140,1005)(1410,1545)
\path(1383.167,1424.252)(1410.000,1545.000)(1329.502,1451.085)
\path(1140,1005)(1590,1905) \path(1140,1005)(690,1905)
\path(1140,1005)(690,1905) \path(1140,1005)(870,1545)
\path(1140,1005)(870,1545)
\whiten\path(950.498,1451.085)(870.000,1545.000)(896.833,1424.252)(939.765,1405.469)(950.498,1451.085)
\path(3030,1005)(3300,1545)
\path(3273.167,1424.252)(3300.000,1545.000)(3219.502,1451.085)
\path(3030,1005)(3480,1905) \path(2580,105)(2850,645)
\path(2823.167,524.252)(2850.000,645.000)(2769.502,551.085)
\path(2580,105)(3030,1005) \path(3480,105)(3030,1005)
\path(3480,105)(3030,1005) \path(3480,105)(3210,645)
\path(3480,105)(3210,645)
\whiten\path(3290.498,551.085)(3210.000,645.000)(3236.833,524.252)(3279.765,505.469)(3290.498,551.085)
\put(3030,1005){\blacken\ellipse{46}{46}}
\put(3030,1005){\ellipse{46}{46}}
\put(2670,1860){\makebox(0,0)[lb]{\smash{{\SetFigFont{12}{14.4}{\rmdefault}{\mddefault}{\updefault}$z_{k-1}$}}}}
\put(3570,1860){\makebox(0,0)[lb]{\smash{{\SetFigFont{12}{14.4}{\rmdefault}{\mddefault}{\updefault}$z_k$}}}}
\put(3120,960){\makebox(0,0)[lb]{\smash{{\SetFigFont{12}{14.4}{\rmdefault}{\mddefault}{\updefault}$y_k$}}}}
\put(2670,60){\makebox(0,0)[lb]{\smash{{\SetFigFont{12}{14.4}{\rmdefault}{\mddefault}{\updefault}$x_k$}}}}
\put(3570,60){\makebox(0,0)[lb]{\smash{{\SetFigFont{12}{14.4}{\rmdefault}{\mddefault}{\updefault}$x_{k+1}$}}}}
\path(3480,105)(3480,1905) \path(3480,105)(3480,1905)
\path(3480,105)(3480,1140) \path(3480,105)(3480,1140)
\whiten\path(3510.000,1020.000)(3480.000,1140.000)(3450.000,1020.000)(3480.000,1056.000)(3510.000,1020.000)
\put(3570,1005){\makebox(0,0)[lb]{\smash{{\SetFigFont{10}{12.0}{\rmdefault}{\mddefault}{\updefault}\makebox[0.8\width][s]{$(\gamma_{n-1}-\gamma_n)$}}}}}
\put(2455,1410){\makebox(0,0)[lb]{\smash{{\SetFigFont{10}{12.0}{\rmdefault}{\mddefault}{\updefault}\makebox[0.8\width][s]{$\gamma_{n-1}$}}}}}
\path(3030,1005)(2580,1905) \path(3030,1005)(2580,1905)
\path(3030,1005)(2760,1545) \path(3030,1005)(2760,1545)
\whiten\path(2840.498,1451.085)(2760.000,1545.000)(2786.833,1424.252)(2829.765,1405.469)(2840.498,1451.085)
\put(735,1410){\makebox(0,0)[lb]{\smash{{\SetFigFont{10}{12.0}{\rmdefault}{\mddefault}{\updefault}$\gamma_n$}}}}
\put(1015,510){\makebox(0,0)[lb]{\smash{{\SetFigFont{10}{12.0}{\rmdefault}{\mddefault}{\updefault}\makebox[0.8\width][s]{$\gamma_{n-1}$}}}}}
\put(3030,510){\makebox(0,0)[lb]{\smash{{\SetFigFont{10}{12.0}{\rmdefault}{\mddefault}{\updefault}$\gamma_n$}}}}
\path(690,105)(690,1905) \path(690,105)(690,1905)
\path(690,105)(690,1140) \path(690,105)(690,1140)
\whiten\path(720.000,1020.000)(690.000,1140.000)(660.000,1020.000)(690.000,1056.000)(720.000,1020.000)
\put(2130,915){\makebox(0,0)[lb]{\smash{{\SetFigFont{14}{16.8}{\rmdefault}{\mddefault}{\updefault}=}}}}
\put(1970,-250){\makebox(0,0)[lb]{\smash{{\SetFigFont{14}{16.8}{\rmdefault}{\mddefault}{\updefault}fig.
3a}}}}
\end{picture}
} \hspace{\fill} \\ [17pt]

\hspace{31mm} \setlength{\unitlength}{0.00087489in}
\begingroup\makeatletter\ifx\SetFigFont\undefined%
\gdef\SetFigFont#1#2#3#4#5{%
  \reset@font\fontsize{#1}{#2pt}%
  \fontfamily{#3}\fontseries{#4}\fontshape{#5}%
  \selectfont}%
\fi\endgroup%
{\renewcommand{\dashlinestretch}{30}
\begin{picture}(3905,1980)(0,-10)
\put(555,1005){\makebox(0,0)[rb]{\smash{{\SetFigFont{10}{12.0}{\rmdefault}{\mddefault}{\updefault}\makebox[0.8\width][s]{$(\gamma_{n-1}-\gamma_n)$}}}}}
\path(1140,1005)(690,1905) \path(1140,1005)(690,1905)
\path(1140,1005)(870,1545) \path(1140,1005)(870,1545)
\whiten\path(950.498,1451.085)(870.000,1545.000)(896.833,1424.252)(939.765,1405.469)(950.498,1451.085)
\path(2580,105)(2850,645)
\path(2823.167,524.252)(2850.000,645.000)(2769.502,551.085)
\path(2580,105)(3030,1005) \path(3480,105)(3030,1005)
\path(3480,105)(3030,1005) \path(3480,105)(3210,645)
\path(3480,105)(3210,645)
\whiten\path(3290.498,551.085)(3210.000,645.000)(3236.833,524.252)(3279.765,505.469)(3290.498,551.085)
\put(3030,1005){\blacken\ellipse{46}{46}}
\put(3030,1005){\ellipse{46}{46}}
\put(2455,1410){\makebox(0,0)[lb]{\smash{{\SetFigFont{10}{12.0}{\rmdefault}{\mddefault}{\updefault}$\gamma_{n-1}$}}}}
\path(3030,1005)(2580,1905) \path(3030,1005)(2580,1905)
\path(3030,1005)(2760,1545) \path(3030,1005)(2760,1545)
\whiten\path(2840.498,1451.085)(2760.000,1545.000)(2786.833,1424.252)(2829.765,1405.469)(2840.498,1451.085)
\put(735,1410){\makebox(0,0)[lb]{\smash{{\SetFigFont{10}{12.0}{\rmdefault}{\mddefault}{\updefault}$\gamma_n$}}}}
\put(1015,510){\makebox(0,0)[lb]{\smash{{\SetFigFont{10}{12.0}{\rmdefault}{\mddefault}{\updefault}$\gamma_{n-1}$}}}}
\put(3030,510){\makebox(0,0)[lb]{\smash{{\SetFigFont{10}{12.0}{\rmdefault}{\mddefault}{\updefault}$\gamma_n$}}}}
\path(690,105)(960,645)
\path(933.167,524.252)(960.000,645.000)(879.502,551.085)
\path(690,105)(1140,1005) \path(1590,105)(1140,1005)
\path(1590,105)(1140,1005) \path(1590,105)(1320,645)
\path(1590,105)(1320,645)
\whiten\path(1400.498,551.085)(1320.000,645.000)(1346.833,524.252)(1389.765,505.469)(1400.498,551.085)
\put(1140,1005){\blacken\ellipse{46}{46}}
\put(1140,1005){\ellipse{46}{46}}
\put(780,1860){\makebox(0,0)[lb]{\smash{{\SetFigFont{12}{14.4}{\rmdefault}{\mddefault}{\updefault}$z_{N-n}$}}}}
\put(2670,1860){\makebox(0,0)[lb]{\smash{{\SetFigFont{12}{14.4}{\rmdefault}{\mddefault}{\updefault}$z_{N-n}$}}}}
\put(1230,960){\makebox(0,0)[lb]{\smash{{\SetFigFont{12}{14.4}{\rmdefault}{\mddefault}{\updefault}$y_{N-n+1}$}}}}
\put(3120,960){\makebox(0,0)[lb]{\smash{{\SetFigFont{12}{14.4}{\rmdefault}{\mddefault}{\updefault}$y_{N-n+1}$}}}}
\put(2670,60){\makebox(0,0)[lb]{\smash{{\SetFigFont{12}{14.4}{\rmdefault}{\mddefault}{\updefault}$x_{N-n+1}$}}}}
\put(3570,60){\makebox(0,0)[lb]{\smash{{\SetFigFont{12}{14.4}{\rmdefault}{\mddefault}{\updefault}$x_{N-n+2}$}}}}
\put(1680,60){\makebox(0,0)[lb]{\smash{{\SetFigFont{12}{14.4}{\rmdefault}{\mddefault}{\updefault}$x_{N-n+2}$}}}}
\put(780,60){\makebox(0,0)[lb]{\smash{{\SetFigFont{12}{14.4}{\rmdefault}{\mddefault}{\updefault}$x_{N-n+1}$}}}}
\path(690,105)(690,1905) \path(690,105)(690,1905)
\path(690,105)(690,1140) \path(690,105)(690,1140)
\whiten\path(720.000,1020.000)(690.000,1140.000)(660.000,1020.000)(690.000,1056.000)(720.000,1020.000)
\put(2030,915){\makebox(0,0)[lb]{\smash{{\SetFigFont{14}{16.8}{\rmdefault}{\mddefault}{\updefault}=}}}}
\put(1870,-250){\makebox(0,0)[lb]{\smash{{\SetFigFont{14}{14.4}{\rmdefault}{\mddefault}{\updefault}fig.
3b}}}}
\end{picture}
} \\ [17pt]

\hspace{35mm} \setlength{\unitlength}{0.00087489in}
\begingroup\makeatletter\ifx\SetFigFont\undefined%
\gdef\SetFigFont#1#2#3#4#5{%
  \reset@font\fontsize{#1}{#2pt}%
  \fontfamily{#3}\fontseries{#4}\fontshape{#5}%
  \selectfont}%
\fi\endgroup%
{\renewcommand{\dashlinestretch}{30}
\begin{picture}(4411,1980)(0,-10)
\path(510,105)(780,645)
\path(753.167,524.252)(780.000,645.000)(699.502,551.085)
\path(510,105)(960,1005) \path(1410,105)(960,1005)
\path(1410,105)(960,1005) \path(1410,105)(1140,645)
\path(1410,105)(1140,645)
\whiten\path(1220.498,551.085)(1140.000,645.000)(1166.833,524.252)(1209.765,505.469)(1220.498,551.085)
\put(960,1005){\blacken\ellipse{46}{46}}
\put(960,1005){\ellipse{46}{46}} \path(960,1005)(1230,1545)
\path(1203.167,1424.252)(1230.000,1545.000)(1149.502,1451.085)
\path(960,1005)(1410,1905)
\put(555,1410){\makebox(0,0)[lb]{\smash{{\SetFigFont{10}{12.0}{\rmdefault}{\mddefault}{\updefault}$\gamma_n$}}}}
\put(835,510){\makebox(0,0)[lb]{\smash{{\SetFigFont{10}{12.0}{\rmdefault}{\mddefault}{\updefault}$\gamma_{n-1}$}}}}
\path(960,1005)(690,1545) \path(960,1005)(690,1545)
\whiten\path(770.498,1451.085)(690.000,1545.000)(716.833,1424.252)(770.498,1451.085)
\path(510,105)(240,645) \path(510,105)(240,645)
\whiten\path(320.498,551.085)(240.000,645.000)(266.833,524.252)(320.498,551.085)
\put(-65,510){\makebox(0,0)[lb]{\smash{{\SetFigFont{10}{12.0}{\rmdefault}{\mddefault}{\updefault}$\gamma_{n-1}$}}}}
\path(3255,1005)(3525,1545)
\path(3498.167,1424.252)(3525.000,1545.000)(3444.502,1451.085)
\path(3255,1005)(3705,1905) \path(2805,105)(3075,645)
\path(3048.167,524.252)(3075.000,645.000)(2994.502,551.085)
\path(2805,105)(3255,1005) \path(3705,105)(3255,1005)
\path(3705,105)(3255,1005) \path(3705,105)(3435,645)
\path(3705,105)(3435,645)
\whiten\path(3515.498,551.085)(3435.000,645.000)(3461.833,524.252)(3504.765,505.469)(3515.498,551.085)
\put(3255,1005){\blacken\ellipse{46}{46}}
\put(3255,1005){\ellipse{46}{46}} \path(3705,105)(3705,1905)
\path(3705,105)(3705,1905) \path(3705,105)(3705,1140)
\path(3705,105)(3705,1140)
\whiten\path(3735.000,1020.000)(3705.000,1140.000)(3675.000,1020.000)(3705.000,1056.000)(3735.000,1020.000)
\put(3795,1005){\makebox(0,0)[lb]{\smash{{\SetFigFont{10}{12.0}{\rmdefault}{\mddefault}{\updefault}\makebox[0.8\width][s]{$(\gamma_{n-1}-\gamma_n)$}}}}}
\put(2680,1410){\makebox(0,0)[lb]{\smash{{\SetFigFont{10}{12.0}{\rmdefault}{\mddefault}{\updefault}$\gamma_{n-1}$}}}}
\put(3255,510){\makebox(0,0)[lb]{\smash{{\SetFigFont{10}{12.0}{\rmdefault}{\mddefault}{\updefault}$\gamma_n$}}}}
\path(3255,1005)(2985,1545) \path(3255,1005)(2985,1545)
\whiten\path(3065.498,1451.085)(2985.000,1545.000)(3011.833,1424.252)(3065.498,1451.085)
\path(2805,105)(2535,645) \path(2805,105)(2535,645)
\whiten\path(2615.498,551.085)(2535.000,645.000)(2561.833,524.252)(2615.498,551.085)
\put(2355,510){\makebox(0,0)[lb]{\smash{{\SetFigFont{10}{12.0}{\rmdefault}{\mddefault}{\updefault}$\gamma_n$}}}}
\put(600,60){\makebox(0,0)[lb]{\smash{{\SetFigFont{12}{14.4}{\rmdefault}{\mddefault}{\updefault}$x_1$}}}}
\put(1500,60){\makebox(0,0)[lb]{\smash{{\SetFigFont{12}{14.4}{\rmdefault}{\mddefault}{\updefault}$x_2$}}}}
\put(1050,960){\makebox(0,0)[lb]{\smash{{\SetFigFont{12}{14.4}{\rmdefault}{\mddefault}{\updefault}$y_1$}}}}
\put(1500,1860){\makebox(0,0)[lb]{\smash{{\SetFigFont{12}{14.4}{\rmdefault}{\mddefault}{\updefault}$z_1$}}}}
\put(1905,915){\makebox(0,0)[lb]{\smash{{\SetFigFont{14}{16.8}{\rmdefault}{\mddefault}{\updefault}=}}}}
\put(2895,60){\makebox(0,0)[lb]{\smash{{\SetFigFont{12}{14.4}{\rmdefault}{\mddefault}{\updefault}$x_1$}}}}
\put(3795,60){\makebox(0,0)[lb]{\smash{{\SetFigFont{12}{14.4}{\rmdefault}{\mddefault}{\updefault}$x_2$}}}}
\put(3345,960){\makebox(0,0)[lb]{\smash{{\SetFigFont{12}{14.4}{\rmdefault}{\mddefault}{\updefault}$y_1$}}}}
\put(3795,1860){\makebox(0,0)[lb]{\smash{{\SetFigFont{12}{14.4}{\rmdefault}{\mddefault}{\updefault}$z_1$}}}}
\put(1745,-250){\makebox(0,0)[lb]{\smash{{\SetFigFont{14}{16.8}{\rmdefault}{\mddefault}{\updefault}fig.
3c}}}}
\end{picture}
} \\ [17pt]

\noindent{\bfseries Proof.} Integration over $y_k$ in the left
hand side of fig.~3a yields
\begin{equation}
\begin{split} \label{rhsf3a}
 \int\limits_{-\infty}^{+\infty} dy_k I(x_k,y_k) J_{\gamma_{n-1}}(x_{k+1},y_k) I(y_k,z_k) & J_{\gamma_n}(y_k,z_{k-1})
 =e^{\frac{i}{\hbar}(\gamma_{n-1}x_{k+1}-\gamma_nz_{k-1})}\times \\
  \times\int\limits_{-\infty}^{+\infty} dy_k
  \exp\Bigl\{\frac{i}{\hbar}(\gamma_n-\gamma_{n-1})y_k-\frac1{\hbar}(e^{-x_{k+1}}&+e^{-z_k})e^{y_k}
             -\frac1{\hbar}(e^{x_k}+e^{z_{k-1}})e^{-y_k}\Bigr\}= \\
 =2e^{\frac{i}{\hbar}(\gamma_{n-1}x_{k+1}-\gamma_nz_{k-1})} & \left( \frac{e^{x_k}+e^{z_{k-1}}}{e^{-x_{k+1}}+e^{-z_k}}
 \right)^{\frac{i(\gamma_n-\gamma_{n-1})}{2\hbar}} \times \\
 \times K_{\frac{i}{\hbar}(\gamma_n-\gamma_{n-1})} & \left(\frac{2}{\hbar}
                                  \sqrt{(e^{x_k}+e^{z_{k-1}})(e^{-x_{k+1}}+e^{-z_k})}\right),
\end{split}
\end{equation}
where $K_\nu (z)$ is a Macdonald function~\cite{BE2}.
Interchanging $\gamma_{n-1}$ and $\gamma_n$ in~\eqref{rhsf3a} we
obtain the expression for the integral in the right hand side of
fig.~3a
\begin{equation} \label{lhsf3a}
\begin{split}
 2e^{\frac{i}{\hbar}(\gamma_n x_{k+1}-\gamma_{n-1}z_{k-1})}&\left( \frac{e^{x_k}+e^{z_{k-1}}}{e^{-x_{k+1}}+e^{-z_k}}
 \right)^{-\frac{i(\gamma_n-\gamma_{n-1})}{2\hbar}} \times \\
 \times K_{\frac{i}{\hbar}(\gamma_n-\gamma_{n-1})} & \left(\frac{2}{\hbar}
                                  \sqrt{(e^{x_k}+e^{z_{k-1}})(e^{-x_{k+1}}+e^{-z_k})}\right).
\end{split}
\end{equation}
The ratio of~\eqref{rhsf3a} and \eqref{lhsf3a} is exactly equal to
$Y_{\gamma_{n-1}-\gamma_n}(x_{k+1},z_k)Y_{\gamma_{n-1}-\gamma_n}^{-1}(x_k,z_{k-1})$. \\

The equalities shown in the fig.~3b and 3c can be proved analogously. \qed \\

Let us continue the proof of the theorem~\ref{Th_psi}. It is shown
in fig.~4. The left hand side of~\eqref{LgammaLgamma} after
application of fig.~3c is reflected in this diagram. Then, using
the fig.~3a, one can move the vertical line picturing the function
$Y_{\gamma_{n-1}-\gamma_n}(x_j,z_{j-1})$ to the right, as shown in
the figure. When this line has arrived to the right one can apply
the fig. 3b. In each step the parameters $\gamma_{n-1}$ and
$\gamma_n$ are interchanged, and,
eventually, one has the right hand side of~\eqref{LgammaLgamma}. \qed \\

\hspace{\fill}\setlength{\unitlength}{0.00087489in}
\begingroup\makeatletter\ifx\SetFigFont\undefined%
\gdef\SetFigFont#1#2#3#4#5{%
  \reset@font\fontsize{#1}{#2pt}%
  \fontfamily{#3}\fontseries{#4}\fontshape{#5}%
  \selectfont}%
\fi\endgroup%
{\renewcommand{\dashlinestretch}{30}
\begin{picture}(5885,2497)(0,-10)
\path(1320,622)(870,1522) \path(1320,622)(870,1522)
\path(1320,622)(1050,1162) \path(1320,622)(1050,1162)
\whiten\path(1130.498,1068.085)(1050.000,1162.000)(1076.833,1041.252)(1119.765,1022.469)(1130.498,1068.085)
\path(420,622)(690,1162)
\path(663.167,1041.252)(690.000,1162.000)(609.502,1068.085)
\path(420,622)(870,1522)
\put(1655,1027){\makebox(0,0)[lb]{\smash{{\SetFigFont{10}{12.0}{\rmdefault}{\mddefault}{\updefault}$\gamma_{n-1}$}}}}
\put(05,1027){\makebox(0,0)[lb]{\smash{{\SetFigFont{10}{12.0}{\rmdefault}{\mddefault}{\updefault}$\gamma_n$}}}}
\put(905,1027){\makebox(0,0)[lb]{\smash{{\SetFigFont{10}{12.0}{\rmdefault}{\mddefault}{\updefault}$\gamma_n$}}}}
\put(1410,2377){\makebox(0,0)[lb]{\smash{{\SetFigFont{12}{14.4}{\rmdefault}{\mddefault}{\updefault}$z_1$}}}}
\put(1350,1972){\makebox(0,0)[lb]{\smash{{\SetFigFont{10}{12.0}{\rmdefault}{\mddefault}{\updefault}$\gamma_n$}}}}
\put(295,1972){\makebox(0,0)[lb]{\smash{{\SetFigFont{10}{12.0}{\rmdefault}{\mddefault}{\updefault}$\gamma_{n-1}$}}}}
\path(4155,1522)(3705,2422) \path(4155,1522)(3705,2422)
\path(4155,1522)(3885,2062) \path(4155,1522)(3885,2062)
\whiten\path(3965.498,1968.085)(3885.000,2062.000)(3911.833,1941.252)(3954.765,1922.469)(3965.498,1968.085)
\path(3705,622)(3975,1162)
\path(3948.167,1041.252)(3975.000,1162.000)(3894.502,1068.085)
\path(3705,622)(4155,1522)
\put(4040,1027){\makebox(0,0)[lb]{\smash{{\SetFigFont{10}{12.0}{\rmdefault}{\mddefault}{\updefault}$\gamma_{n-1}$}}}}
\put(4940,1027){\makebox(0,0)[lb]{\smash{{\SetFigFont{10}{12.0}{\rmdefault}{\mddefault}{\updefault}$\gamma_{n-1}$}}}}
\put(3705,1972){\makebox(0,0)[lb]{\smash{{\SetFigFont{10}{12.0}{\rmdefault}{\mddefault}{\updefault}$\gamma_n$}}}}
\put(4610,1972){\makebox(0,0)[lb]{\smash{{\SetFigFont{10}{12.0}{\rmdefault}{\mddefault}{\updefault}$\gamma_n$}}}}

\put(1020,1702){\makebox(0,0)[lb]{\smash{{\SetFigFont{10}{12.0}{\rmdefault}{\mddefault}{\updefault}\makebox[0.7\width][s]{$(\gamma_{n-1}-\gamma_n)$}}}}}
\put(870,1522){\blacken\ellipse{46}{46}}
\put(870,1522){\ellipse{46}{46}}
\put(1770,1522){\blacken\ellipse{46}{46}}
\put(1770,1522){\ellipse{46}{46}}
\put(4155,1522){\blacken\ellipse{46}{46}}
\put(4155,1522){\ellipse{46}{46}}
\put(5055,1522){\blacken\ellipse{46}{46}}
\put(5055,1522){\ellipse{46}{46}} \path(870,1522)(600,2062)
\path(870,1522)(600,2062)
\whiten\path(680.498,1968.085)(600.000,2062.000)(626.833,1941.252)(680.498,1968.085)
\path(420,622)(150,1162) \path(420,622)(150,1162)
\whiten\path(230.498,1068.085)(150.000,1162.000)(176.833,1041.252)(230.498,1068.085)
\path(1320,1747)(1320,2422) \path(1320,1747)(1320,2422)
\path(1320,622)(1320,1657) \path(1320,622)(1320,1657)
\whiten\path(1350.000,1537.000)(1320.000,1657.000)(1290.000,1537.000)(1320.000,1573.000)(1350.000,1537.000)
\path(870,1522)(1140,2062)
\path(1113.167,1941.252)(1140.000,2062.000)(1059.502,1968.085)
\path(870,1522)(1320,2422) \path(1635,1792)(1320,2422)
\path(1635,1792)(1320,2422) \path(1635,1792)(1500,2062)
\path(1635,1792)(1500,2062)
\whiten\path(1580.498,1968.085)(1500.000,2062.000)(1526.833,1941.252)(1569.765,1922.469)(1580.498,1968.085)
\path(2220,622)(1710,1642) \path(2220,622)(1710,1642)
\path(2220,622)(1950,1162) \path(2220,622)(1950,1162)
\whiten\path(2030.498,1068.085)(1950.000,1162.000)(1976.833,1041.252)(2019.765,1022.469)(2030.498,1068.085)
\path(1320,622)(1590,1162)
\path(1563.167,1041.252)(1590.000,1162.000)(1509.502,1068.085)
\path(1320,622)(1770,1522) \path(1770,1522)(2040,2062)
\path(2013.167,1941.252)(2040.000,2062.000)(1959.502,1968.085)
\path(1770,1522)(2220,2422) \path(1770,1522)(2220,2422)
\path(5505,622)(5055,1522) \path(5505,622)(5055,1522)
\path(5505,622)(5235,1162) \path(5505,622)(5235,1162)
\whiten\path(5315.498,1068.085)(5235.000,1162.000)(5261.833,1041.252)(5304.765,1022.469)(5315.498,1068.085)
\path(4605,622)(4875,1162)
\path(4848.167,1041.252)(4875.000,1162.000)(4794.502,1068.085)
\path(4605,622)(5055,1522) \path(5055,1522)(4605,2422)
\path(5055,1522)(4605,2422) \path(5055,1522)(4785,2062)
\path(5055,1522)(4785,2062)
\whiten\path(4865.498,1968.085)(4785.000,2062.000)(4811.833,1941.252)(4854.765,1922.469)(4865.498,1968.085)
\path(4605,622)(4155,1522) \path(4605,622)(4155,1522)
\path(4605,622)(4335,1162) \path(4605,622)(4335,1162)
\whiten\path(4415.498,1068.085)(4335.000,1162.000)(4361.833,1041.252)(4404.765,1022.469)(4415.498,1068.085)
\path(4155,1522)(4425,2062)
\path(4398.167,1941.252)(4425.000,2062.000)(4344.502,1968.085)
\path(4155,1522)(4605,2422)
\dashline{60.000}(1365,1297)(4515,1297)
\blacken\path(4395.000,1267.000)(4515.000,1297.000)(4395.000,1327.000)(4395.000,1267.000)
\put(960,1477){\makebox(0,0)[lb]{\smash{{\SetFigFont{12}{14.4}{\rmdefault}{\mddefault}{\updefault}$y_1$}}}}
\put(1860,1477){\makebox(0,0)[lb]{\smash{{\SetFigFont{12}{14.4}{\rmdefault}{\mddefault}{\updefault}$y_2$}}}}
\put(510,577){\makebox(0,0)[lb]{\smash{{\SetFigFont{12}{14.4}{\rmdefault}{\mddefault}{\updefault}$x_1$}}}}
\put(2310,577){\makebox(0,0)[lb]{\smash{{\SetFigFont{12}{14.4}{\rmdefault}{\mddefault}{\updefault}$x_3$}}}}
\put(1410,577){\makebox(0,0)[lb]{\smash{{\SetFigFont{12}{14.4}{\rmdefault}{\mddefault}{\updefault}$x_2$}}}}
\put(2310,2377){\makebox(0,0)[lb]{\smash{{\SetFigFont{12}{14.4}{\rmdefault}{\mddefault}{\updefault}$z_2$}}}}
\put(2895,892){\makebox(0,0)[lb]{\smash{{\SetFigFont{20}{24.0}{\rmdefault}{\mddefault}{\updefault}...}}}}
\put(2715,82){\makebox(0,0)[lb]{\smash{{\SetFigFont{14}{16.8}{\rmdefault}{\mddefault}{\updefault}fig.
4}}}}
\put(4695,577){\makebox(0,0)[lb]{\smash{{\SetFigFont{12}{14.4}{\rmdefault}{\mddefault}{\updefault}$x_{N-n+1}$}}}}
\put(4245,1477){\makebox(0,0)[lb]{\smash{{\SetFigFont{12}{14.4}{\rmdefault}{\mddefault}{\updefault}$y_{N-n}$}}}}
\put(5145,1477){\makebox(0,0)[lb]{\smash{{\SetFigFont{12}{14.4}{\rmdefault}{\mddefault}{\updefault}$y_{N-n+1}$}}}}
\put(4695,2377){\makebox(0,0)[lb]{\smash{{\SetFigFont{12}{14.4}{\rmdefault}{\mddefault}{\updefault}$z_{N-n}$}}}}
\put(3795,2377){\makebox(0,0)[lb]{\smash{{\SetFigFont{12}{14.4}{\rmdefault}{\mddefault}{\updefault}$z_{N-n-1}$}}}}
\put(3795,577){\makebox(0,0)[lb]{\smash{{\SetFigFont{12}{14.4}{\rmdefault}{\mddefault}{\updefault}$x_{N-n}$}}}}
\put(5550,577){\makebox(0,0)[lb]{\smash{{\SetFigFont{12}{14.4}{\rmdefault}{\mddefault}{\updefault}$x_{N-n+2}$}}}}
\end{picture}
}
\hspace{\fill} \\

Substituting the expression~\eqref{Lambf} for the operator $\Lambda_N(\gamma_N)$ with the kernel~\eqref{Lamb} to~\eqref{psi} one obtains the recurrent formula
\begin{equation} \label{psi_rec}
 \begin{split}
 \psi_{\gamma_1,\ldots,\gamma_N}(x_1,\ldots,x_N)=\int\limits_{\mathbb R^{N-1}} dy_1\ldots dy_{N-1}\,
  \psi_{\gamma_1,\ldots,\gamma_{N-1}}(y_1,\ldots,y_{N-1}) \times \\
 \times\exp\Bigl\{\frac{i}{\hbar}\gamma_N(\sum\limits_{n=1}^N x_n-\sum\limits_{n=1}^{N-1}y_n)
        -\frac{1}{\hbar}\sum\limits_{n=1}^{N-1}(e^{y_n-x_{n+1}}+e^{x_n-y_n})\Bigr\}.
 \end{split}
\end{equation}
The consecutive applications of this formula allow to derive the following integral representation for the
eigenfunctions of open Toda chain model
\begin{align}
 \psi_{\gamma}(z_{N1},\ldots,z_{NN})=\int\limits_{\mathbb R^{\frac{N(N-1)}2}} \prod_{n=1}^{N-1}\prod_{j=1}^{n}dz_{nj}\,
  &\exp\biggl\{\frac{i}{\hbar}\Bigl[\gamma_N\sum\limits_{j=1}^N z_{Nj}
    +\sum_{n=1}^{N-1}(\gamma_n-\gamma_{n+1})\sum_{j=1}^n z_{nj}\Bigr]- \notag \\
    -&\frac{1}{\hbar}\sum_{n=1}^N\sum_{j=1}^{n-1}\bigl(e^{z_{nj}-z_{n-1,j}}
    +e^{z_{n-1,j}-z_{n,j+1}}\bigr)\biggr\}.\label{psi_GG}
\end{align}
This is a Gauss-Givental representation~\cite{Kharchev_GG},
\cite{Givental} of the Toda chain transition function.

\section{Integration measure}
\label{Int_meas}

In this section we shall check the orthogonality condition using
the diagram technique. The normalization function which appears in
this calculation coincides exactly with the Sklyanin integration
measure using in the SoV method for the periodic Toda
chain model~\cite{Sklyanin},
\cite{Kharchev_P},\cite{Kharchev_O}, \cite{Kharchev_OP}. \\

\begin{theorem} \label{orthog_meas}
 The functions $\psi_{\gamma}(x)$ defined by the formula~\eqref{psi} satisfy to the
orthogonality condition
\begin{equation} \label{orthog}
 \int\limits_{\mathbb R^N} dx \overline{\psi_{\gamma}(x)}\psi_{\gamma'}(x)=\mu^{-1}(\gamma)\delta_{SYM}(\gamma,\gamma'),
\end{equation}
where
\begin{align*}
 \delta_{SYM}(\gamma,\gamma')=\dfrac1{N!}\sum\limits_{\sigma\in S_N}\prod\limits_{i=1}^N\delta(\gamma_i-\gamma'_{\sigma(i)})
\end{align*}
is a symmetrized delta function and $\mu(\gamma)$ is the Sklyanin measure
 \begin{equation} \label{measure}
  \mu(\gamma)=\frac{(2\pi\hbar)^{-N}}{N!}\prod_{k<m}
  \left[\Gamma\Bigl(\frac{\gamma_m-\gamma_k}{i\hbar}\Bigr)
        \Gamma\Bigl(\frac{\gamma_k-\gamma_m}{i\hbar}\Bigr)\right]^{-1}.
 \end{equation}
The both sides of equality~\eqref{orthog} are understand as distributions with arguments $\gamma'_1,\ldots,\gamma'_N$, depending on the pairwise different parameters $\gamma_1,\ldots,\gamma_N$.
\end{theorem}

Since the integrals of motion $H_k$ are Hermitian operators, the set of the functions
$\psi_{\gamma}(x)$ is complete \cite{Gelfand4}. This means that
any function $f(x)$ belonging to the Hilbert space $L^2(\mathbb
R^N)$ can be represented as an integral
\begin{equation} \label{f_psi}
 f(x)=\int\limits_{\mathbb R^N} \psi_{\gamma}(x) g(\gamma)\mu(\gamma)\,d\gamma
\end{equation}
for some summable function $g(\gamma)$.
As a consequence we have the completeness condition
\begin{equation} \label{compl}
 \int\limits_{\mathbb R^N} d\gamma\,\mu(\gamma)\overline{\psi_{\gamma}(x)}\psi_{\gamma}(x')=\prod_{i=1}^N\delta(x_i-x'_i),
\end{equation}.

{\bfseries Proof of the theorem~\ref{orthog_meas}.} First of all,
we need to obtain a diagram representation of
$\overline{\psi_{\gamma}(x)}$ in order to calculate the integral
in the left hand side of~\eqref{orthog} using the diagram
technique. Let us to consider the diagram for $\psi_{\gamma}(x)$,
which are shown in fig.~2 for $N=5$ and in fig.~6a for $N=3$, and
to implement the following steps.

{\bfseries First step.} The imaginary unit is contained only in the
functions $J$ and $Z$. To reduce the conjugation of whole function
$\psi_{\gamma}(x)$ to the conjugation of the functions $Z$ we
decompose $J_u(x_{k+1},y_k)$ into the product
$Z_u^{-1}(y_k)I(y_k,x_{k+1})Z_u(x_{k+1})$ (fig.~5a). This
corresponds to the transition from the fig.~6a to the fig.~6b.

{\bfseries Second step.} We replace all $Z_u(x_k)$ by
$Z_{u}^{-1}(x_k)$ (fig.~5b) implementing the complex conjugation
and arrive to the fig.~6c, in which the function
$\overline{\psi_{\gamma}(x)}$ is pictured. \\ [8pt]

\noindent\hspace{\fill} \setlength{\unitlength}{0.00087489in}
\begingroup\makeatletter\ifx\SetFigFont\undefined%
\gdef\SetFigFont#1#2#3#4#5{%
  \reset@font\fontsize{#1}{#2pt}%
  \fontfamily{#3}\fontseries{#4}\fontshape{#5}%
  \selectfont}%
\fi\endgroup%
{\renewcommand{\dashlinestretch}{30}
\begin{picture}(1636,1426)(0,-10)
\put(555,388){\makebox(0,0)[lb]{\smash{{\SetFigFont{12}{14.4}{\rmdefault}{\mddefault}{\updefault}$x_{k+1}$}}}}
\path(915,1333)(1185,793)
\path(1104.502,886.915)(1185.000,793.000)(1158.167,913.748)
\path(915,1333)(1365,433)
\put(1455,388){\makebox(0,0)[lb]{\smash{{\SetFigFont{12}{14.4}{\rmdefault}{\mddefault}{\updefault}$x_{k+1}$}}}}
\path(1050,1063)(1005,1153)
\whiten\path(1085.498,1059.085)(1005.000,1153.000)(1031.833,1032.252)(1085.498,1059.085)
\path(1230,703)(1185,793)
\whiten\path(1265.498,699.085)(1185.000,793.000)(1211.833,672.252)(1265.498,699.085)
\put(870,1018){\makebox(0,0)[lb]{\smash{{\SetFigFont{10}{14.4}{\rmdefault}{\mddefault}{\updefault}$u$}}}}
\put(1005,658){\makebox(0,0)[lb]{\smash{{\SetFigFont{10}{14.4}{\rmdefault}{\mddefault}{\updefault}$u$}}}}
\put(1005,1288){\makebox(0,0)[lb]{\smash{{\SetFigFont{12}{14.4}{\rmdefault}{\mddefault}{\updefault}$y_k$}}}}
\path(465,433)(15,1333) \path(465,433)(15,1333)
\path(465,433)(195,973) \path(465,433)(195,973)
\whiten\path(275.498,879.085)(195.000,973.000)(221.833,852.252)(264.765,833.469)(275.498,879.085)
\path(510,883)(825,883)
\blacken\path(705.000,853.000)(825.000,883.000)(705.000,913.000)(741.000,883.000)(705.000,853.000)
\put(15,883){\makebox(0,0)[lb]{\smash{{\SetFigFont{10}{14.4}{\rmdefault}{\mddefault}{\updefault}$u$}}}}
\put(105,1288){\makebox(0,0)[lb]{\smash{{\SetFigFont{12}{14.4}{\rmdefault}{\mddefault}{\updefault}$y_k$}}}}
\put(455,73){\makebox(0,0)[lb]{\smash{{\SetFigFont{14}{14.4}{\rmdefault}{\mddefault}{\updefault}fig.
5a}}}}
\end{picture}
} \hspace{\fill} \setlength{\unitlength}{0.00087489in}
\begingroup\makeatletter\ifx\SetFigFont\undefined%
\gdef\SetFigFont#1#2#3#4#5{%
  \reset@font\fontsize{#1}{#2pt}%
  \fontfamily{#3}\fontseries{#4}\fontshape{#5}%
  \selectfont}%
\fi\endgroup%
{\renewcommand{\dashlinestretch}{30}
\begin{picture}(1197,1000)(0,-10)
\path(150,973)(330,613) \path(1185,613)(1005,973)
\put(870,748){\makebox(0,0)[lb]{\smash{{\SetFigFont{10}{14.4}{\rmdefault}{\mddefault}{\updefault}$u$}}}}
\path(420,838)(735,838)
\blacken\path(615.000,808.000)(735.000,838.000)(615.000,868.000)(651.000,838.000)(615.000,808.000)
\put(15,748){\makebox(0,0)[lb]{\smash{{\SetFigFont{10}{14.4}{\rmdefault}{\mddefault}{\updefault}$u$}}}}
\put(365,73){\makebox(0,0)[lb]{\smash{{\SetFigFont{14}{14.4}{\rmdefault}{\mddefault}{\updefault}fig.
5b}}}} \path(240,793)(195,883)
\whiten\path(275.498,789.085)(195.000,883.000)(221.833,762.252)(275.498,789.085)
\path(1095,793)(1140,703)
\whiten\path(1059.502,796.915)(1140.000,703.000)(1113.167,823.748)(1059.502,796.915)
\end{picture}
} \hspace{\fill} \\ [10pt]

{\bfseries Third step.} Since the functions $Z_{u}^{-1}(x_k)$ are
attached to only one point, namely to $x_k$, one can turn these
functions in the manner shown in fig.~5c and fig.~5d. It means
that we can represent $\overline{\psi_{\gamma}(x)}$ by the
fig.~6d.

{\bfseries Fourth step.} Now we replace the product
$Z_u^{-1}(x_k)I(x_k,y_k)Z_u(y_k)$ by $J_u(y_k,x_k)$ (fig.~5e) to
obtain fig.~6e. \\ [8pt]

\noindent\hspace{\fill} \setlength{\unitlength}{0.00087489in}
\begingroup\makeatletter\ifx\SetFigFont\undefined%
\gdef\SetFigFont#1#2#3#4#5{%
  \reset@font\fontsize{#1}{#2pt}%
  \fontfamily{#3}\fontseries{#4}\fontshape{#5}%
  \selectfont}%
\fi\endgroup%
{\renewcommand{\dashlinestretch}{30}
\begin{picture}(1519,1138)(0,-10)
\put(303,1018){\makebox(0,0)[rb]{\smash{{\SetFigFont{12}{14.4}{\rmdefault}{\mddefault}{\updefault}$x_k$}}}}
\put(1248,1041){\blacken\ellipse{46}{46}}
\put(1248,1041){\ellipse{46}{46}} \path(1023,613)(1248,1063)
\put(1338,1018){\makebox(0,0)[lb]{\smash{{\SetFigFont{12}{14.4}{\rmdefault}{\mddefault}{\updefault}$x_k$}}}}
\put(933,748){\makebox(0,0)[lb]{\smash{{\SetFigFont{10}{14.4}{\rmdefault}{\mddefault}{\updefault}$u$}}}}
\put(78,1041){\blacken\ellipse{46}{46}}
\put(78,1041){\ellipse{46}{46}} \path(303,613)(78,1063)
\path(438,838)(753,838)
\blacken\path(633.000,808.000)(753.000,838.000)(633.000,868.000)(669.000,838.000)(633.000,808.000)
\put(123,748){\makebox(0,0)[rb]{\smash{{\SetFigFont{10}{14.4}{\rmdefault}{\mddefault}{\updefault}$u$}}}}
\put(473,73){\makebox(0,0)[lb]{\smash{{\SetFigFont{14}{14.4}{\rmdefault}{\mddefault}{\updefault}fig.
5c}}}} \path(1113,793)(1068,703)
\whiten\path(1094.833,823.748)(1068.000,703.000)(1148.498,796.915)(1094.833,823.748)
\path(213,793)(258,703)
\whiten\path(177.502,796.915)(258.000,703.000)(231.167,823.748)(177.502,796.915)
\end{picture}
} \hspace{\fill} \setlength{\unitlength}{0.00087489in}
\begingroup\makeatletter\ifx\SetFigFont\undefined%
\gdef\SetFigFont#1#2#3#4#5{%
  \reset@font\fontsize{#1}{#2pt}%
  \fontfamily{#3}\fontseries{#4}\fontshape{#5}%
  \selectfont}%
\fi\endgroup%
{\renewcommand{\dashlinestretch}{30}
\begin{picture}(1339,1045)(0,-10)
\put(573,523){\makebox(0,0)[rb]{\smash{{\SetFigFont{12}{14.4}{\rmdefault}{\mddefault}{\updefault}$x_k$}}}}
\put(348,568){\blacken\ellipse{46}{46}}
\put(348,568){\ellipse{46}{46}}
\put(1068,568){\blacken\ellipse{46}{46}}
\put(1068,568){\ellipse{46}{46}} \path(528,793)(843,793)
\blacken\path(723.000,763.000)(843.000,793.000)(723.000,823.000)(759.000,793.000)(723.000,763.000)
\path(348,568)(123,1018) \path(1068,568)(1293,1018)
\put(428,73){\makebox(0,0)[lb]{\smash{{\SetFigFont{14}{14.4}{\rmdefault}{\mddefault}{\updefault}fig.
5d}}}}
\put(123,793){\makebox(0,0)[rb]{\smash{{\SetFigFont{10}{14.4}{\rmdefault}{\mddefault}{\updefault}$u$}}}}
\put(1023,793){\makebox(0,0)[lb]{\smash{{\SetFigFont{10}{14.4}{\rmdefault}{\mddefault}{\updefault}$u$}}}}
\put(1158,523){\makebox(0,0)[lb]{\smash{{\SetFigFont{12}{14.4}{\rmdefault}{\mddefault}{\updefault}$x_k$}}}}
\path(213,838)(258,748)
\whiten\path(177.502,841.915)(258.000,748.000)(231.167,868.748)(177.502,841.915)
\path(1203,838)(1158,748)
\whiten\path(1184.833,868.748)(1158.000,748.000)(1238.498,841.915)(1184.833,868.748)
\end{picture}
} \hspace{\fill} \setlength{\unitlength}{0.00087489in}
\begingroup\makeatletter\ifx\SetFigFont\undefined%
\gdef\SetFigFont#1#2#3#4#5{%
  \reset@font\fontsize{#1}{#2pt}%
  \fontfamily{#3}\fontseries{#4}\fontshape{#5}%
  \selectfont}%
\fi\endgroup%
{\renewcommand{\dashlinestretch}{30}
\begin{picture}(1726,1426)(0,-10)
\path(60,433)(330,973)
\path(303.167,852.252)(330.000,973.000)(249.502,879.085)
\path(60,433)(510,1333) \path(555,883)(870,883)
\blacken\path(750.000,853.000)(870.000,883.000)(750.000,913.000)(786.000,883.000)(750.000,853.000)
\path(195,703)(150,613)
\whiten\path(176.833,733.748)(150.000,613.000)(230.498,706.915)(176.833,733.748)
\path(375,1063)(330,973)
\whiten\path(356.833,1093.748)(330.000,973.000)(410.498,1066.915)(356.833,1093.748)
\path(1455,1333)(1005,433) \path(1455,1333)(1005,433)
\path(1455,1333)(1185,793) \path(1455,1333)(1185,793)
\whiten\path(1211.833,913.748)(1185.000,793.000)(1265.498,886.915)(1254.765,932.531)(1211.833,913.748)
\put(15,658){\makebox(0,0)[lb]{\smash{{\SetFigFont{10}{14.4}{\rmdefault}{\mddefault}{\updefault}$u$}}}}
\put(150,1018){\makebox(0,0)[lb]{\smash{{\SetFigFont{10}{14.4}{\rmdefault}{\mddefault}{\updefault}$u$}}}}
\put(1005,833){\makebox(0,0)[lb]{\smash{{\SetFigFont{10}{14.4}{\rmdefault}{\mddefault}{\updefault}$u$}}}}
\put(600,1288){\makebox(0,0)[lb]{\smash{{\SetFigFont{12}{14.4}{\rmdefault}{\mddefault}{\updefault}$y_k$}}}}
\put(1545,1288){\makebox(0,0)[lb]{\smash{{\SetFigFont{12}{14.4}{\rmdefault}{\mddefault}{\updefault}$y_k$}}}}
\put(150,388){\makebox(0,0)[lb]{\smash{{\SetFigFont{12}{14.4}{\rmdefault}{\mddefault}{\updefault}$x_k$}}}}
\put(1095,388){\makebox(0,0)[lb]{\smash{{\SetFigFont{12}{14.4}{\rmdefault}{\mddefault}{\updefault}$x_k$}}}}
\put(500,73){\makebox(0,0)[lb]{\smash{{\SetFigFont{14}{14.4}{\rmdefault}{\mddefault}{\updefault}fig.
5e}}}}
\end{picture}
} \hspace{\fill} \\ [10pt]

{\bfseries Fifth step.} Finally, reflecting the fig.~6e with respect to a horizontal line
we obtain the fig.~6f. \\ [8pt]

\noindent \setlength{\unitlength}{0.0008in}
\begingroup\makeatletter\ifx\SetFigFont\undefined%
\gdef\SetFigFont#1#2#3#4#5{%
  \reset@font\fontsize{#1}{#2pt}%
  \fontfamily{#3}\fontseries{#4}\fontshape{#5}%
  \selectfont}%
\fi\endgroup%
{\renewcommand{\dashlinestretch}{30}
\begin{picture}(8169,2814)(0,-10)
\path(1182,447)(1452,987)
\path(1425.167,866.252)(1452.000,987.000)(1371.502,893.085)
\path(1182,447)(1632,1347) \path(2082,447)(1632,1347)
\path(2082,447)(1632,1347) \path(2082,447)(1812,987)
\path(2082,447)(1812,987)
\whiten\path(1892.498,893.085)(1812.000,987.000)(1838.833,866.252)(1881.765,847.469)(1892.498,893.085)
\path(1182,447)(732,1347) \path(1182,447)(732,1347)
\path(1182,447)(912,987) \path(1182,447)(912,987)
\whiten\path(992.498,893.085)(912.000,987.000)(938.833,866.252)(981.765,847.469)(992.498,893.085)
\path(3522,1347)(3792,807)
\path(3711.502,900.915)(3792.000,807.000)(3765.167,927.748)
\path(3522,1347)(3972,447) \path(3837,717)(3792,807)
\whiten\path(3872.498,713.085)(3792.000,807.000)(3818.833,686.252)(3872.498,713.085)
\path(3657,1077)(3612,1167)
\whiten\path(3692.498,1073.085)(3612.000,1167.000)(3638.833,1046.252)(3692.498,1073.085)
\path(3972,2247)(4242,1707)
\path(4161.502,1800.915)(4242.000,1707.000)(4215.167,1827.748)
\path(3972,2247)(4422,1347) \path(4287,1617)(4242,1707)
\whiten\path(4322.498,1613.085)(4242.000,1707.000)(4268.833,1586.252)(4322.498,1613.085)
\path(4107,1977)(4062,2067)
\whiten\path(4142.498,1973.085)(4062.000,2067.000)(4088.833,1946.252)(4142.498,1973.085)
\path(4422,1347)(4692,807)
\path(4611.502,900.915)(4692.000,807.000)(4665.167,927.748)
\path(4422,1347)(4872,447) \path(4737,717)(4692,807)
\whiten\path(4772.498,713.085)(4692.000,807.000)(4718.833,686.252)(4772.498,713.085)
\path(4557,1077)(4512,1167)
\whiten\path(4592.498,1073.085)(4512.000,1167.000)(4538.833,1046.252)(4592.498,1073.085)
\path(6762,2247)(6537,2697) \path(6762,2247)(6537,2697)
\path(6537,2697)(6582,2607) \path(6537,2697)(6582,2607)
\whiten\path(6501.502,2700.915)(6582.000,2607.000)(6555.167,2727.748)(6501.502,2700.915)
\path(6312,1347)(6087,1797) \path(6312,1347)(6087,1797)
\path(6087,1797)(6132,1707) \path(6087,1797)(6132,1707)
\whiten\path(6051.502,1800.915)(6132.000,1707.000)(6105.167,1827.748)(6051.502,1800.915)
\path(5862,447)(5637,897) \path(5862,447)(5637,897)
\path(5637,897)(5682,807) \path(5637,897)(5682,807)
\whiten\path(5601.502,900.915)(5682.000,807.000)(5655.167,927.748)(5601.502,900.915)
\path(7212,1347)(6762,2247) \path(6762,2247)(7032,1707)
\path(6951.502,1800.915)(7032.000,1707.000)(7005.167,1827.748)
\path(6762,447)(6312,1347) \path(6312,1347)(6582,807)
\path(6501.502,900.915)(6582.000,807.000)(6555.167,927.748)
\path(7662,447)(7212,1347) \path(7212,1347)(7482,807)
\path(7401.502,900.915)(7482.000,807.000)(7455.167,927.748)
\put(732,1347){\blacken\ellipse{46}{46}}
\put(732,1347){\ellipse{46}{46}}
\put(1182,2247){\blacken\ellipse{46}{46}}
\put(1182,2247){\ellipse{46}{46}}
\put(1632,1347){\blacken\ellipse{46}{46}}
\put(1632,1347){\ellipse{46}{46}}
\put(3522,1347){\blacken\ellipse{46}{46}}
\put(3522,1347){\ellipse{46}{46}}
\put(3972,2247){\blacken\ellipse{46}{46}}
\put(3972,2247){\ellipse{46}{46}}
\put(4422,1347){\blacken\ellipse{46}{46}}
\put(4422,1347){\ellipse{46}{46}}
\put(6312,1347){\blacken\ellipse{46}{46}}
\put(6312,1347){\ellipse{46}{46}}
\put(6762,2247){\blacken\ellipse{46}{46}}
\put(6762,2247){\ellipse{46}{46}}
\put(7212,1347){\blacken\ellipse{46}{46}}
\put(7212,1347){\ellipse{46}{46}} \path(732,1347)(462,1887)
\path(732,1347)(462,1887)
\whiten\path(542.498,1793.085)(462.000,1887.000)(488.833,1766.252)(542.498,1793.085)
\path(1182,2247)(912,2787) \path(1182,2247)(912,2787)
\whiten\path(992.498,2693.085)(912.000,2787.000)(938.833,2666.252)(992.498,2693.085)
\path(282,447)(12,987) \path(282,447)(12,987)
\whiten\path(92.498,893.085)(12.000,987.000)(38.833,866.252)(92.498,893.085)
\path(1632,1347)(1182,2247) \path(1632,1347)(1182,2247)
\path(1632,1347)(1362,1887) \path(1632,1347)(1362,1887)
\whiten\path(1442.498,1793.085)(1362.000,1887.000)(1388.833,1766.252)(1431.765,1747.469)(1442.498,1793.085)
\path(732,1347)(1002,1887)
\path(975.167,1766.252)(1002.000,1887.000)(921.502,1793.085)
\path(732,1347)(1182,2247) \path(282,447)(552,987)
\path(525.167,866.252)(552.000,987.000)(471.502,893.085)
\path(282,447)(732,1347) \path(2172,1527)(2667,1527)
\blacken\path(2547.000,1497.000)(2667.000,1527.000)(2547.000,1557.000)(2583.000,1527.000)(2547.000,1497.000)
\path(4962,1527)(5457,1527)
\blacken\path(5337.000,1497.000)(5457.000,1527.000)(5337.000,1557.000)(5373.000,1527.000)(5337.000,1497.000)
\path(3522,1347)(3252,1887) \path(3522,1347)(3252,1887)
\whiten\path(3332.498,1793.085)(3252.000,1887.000)(3278.833,1766.252)(3332.498,1793.085)
\path(3972,2247)(3702,2787) \path(3972,2247)(3702,2787)
\whiten\path(3782.498,2693.085)(3702.000,2787.000)(3728.833,2666.252)(3782.498,2693.085)
\path(3072,447)(2802,987) \path(3072,447)(2802,987)
\whiten\path(2882.498,893.085)(2802.000,987.000)(2828.833,866.252)(2882.498,893.085)
\path(3522,1347)(3792,1887)
\path(3765.167,1766.252)(3792.000,1887.000)(3711.502,1793.085)
\path(3522,1347)(3972,2247) \path(3072,447)(3342,987)
\path(3315.167,866.252)(3342.000,987.000)(3261.502,893.085)
\path(3072,447)(3522,1347) \path(3972,447)(4242,987)
\path(4215.167,866.252)(4242.000,987.000)(4161.502,893.085)
\path(3972,447)(4422,1347) \path(6312,1347)(6582,1887)
\path(6555.167,1766.252)(6582.000,1887.000)(6501.502,1793.085)
\path(6312,1347)(6762,2247) \path(5862,447)(6132,987)
\path(6105.167,866.252)(6132.000,987.000)(6051.502,893.085)
\path(5862,447)(6312,1347) \path(6762,447)(7032,987)
\path(7005.167,866.252)(7032.000,987.000)(6951.502,893.085)
\path(6762,447)(7212,1347) \path(7662,1527)(8157,1527)
\blacken\path(8037.000,1497.000)(8157.000,1527.000)(8037.000,1557.000)(8073.000,1527.000)(8037.000,1497.000)
\put(822,1302){\makebox(0,0)[lb]{\smash{{\SetFigFont{12}{14.4}{\rmdefault}{\mddefault}{\updefault}$y_1$}}}}
\put(1272,402){\makebox(0,0)[lb]{\smash{{\SetFigFont{12}{14.4}{\rmdefault}{\mddefault}{\updefault}$x_2$}}}}
\put(2172,402){\makebox(0,0)[lb]{\smash{{\SetFigFont{12}{14.4}{\rmdefault}{\mddefault}{\updefault}$x_3$}}}}
\put(1722,1302){\makebox(0,0)[lb]{\smash{{\SetFigFont{12}{14.4}{\rmdefault}{\mddefault}{\updefault}$y_2$}}}}
\put(1272,2202){\makebox(0,0)[lb]{\smash{{\SetFigFont{12}{14.4}{\rmdefault}{\mddefault}{\updefault}$z_3$}}}}
\put(372,402){\makebox(0,0)[lb]{\smash{{\SetFigFont{12}{14.4}{\rmdefault}{\mddefault}{\updefault}$x_1$}}}}
\put(867,87){\makebox(0,0)[lb]{\smash{{\SetFigFont{14}{16.8}{\rmdefault}{\mddefault}{\updefault}fig.
6a}}}}
\put(3612,1302){\makebox(0,0)[lb]{\smash{{\SetFigFont{12}{14.4}{\rmdefault}{\mddefault}{\updefault}$y_1$}}}}
\put(4062,402){\makebox(0,0)[lb]{\smash{{\SetFigFont{12}{14.4}{\rmdefault}{\mddefault}{\updefault}$x_2$}}}}
\put(4962,402){\makebox(0,0)[lb]{\smash{{\SetFigFont{12}{14.4}{\rmdefault}{\mddefault}{\updefault}$x_3$}}}}
\put(4512,1302){\makebox(0,0)[lb]{\smash{{\SetFigFont{12}{14.4}{\rmdefault}{\mddefault}{\updefault}$y_2$}}}}
\put(4062,2202){\makebox(0,0)[lb]{\smash{{\SetFigFont{12}{14.4}{\rmdefault}{\mddefault}{\updefault}$z_1$}}}}
\put(3162,402){\makebox(0,0)[lb]{\smash{{\SetFigFont{12}{14.4}{\rmdefault}{\mddefault}{\updefault}$x_1$}}}}
\put(3657,87){\makebox(0,0)[lb]{\smash{{\SetFigFont{14}{16.8}{\rmdefault}{\mddefault}{\updefault}fig.
6b}}}}
\put(6402,1302){\makebox(0,0)[lb]{\smash{{\SetFigFont{12}{14.4}{\rmdefault}{\mddefault}{\updefault}$y_1$}}}}
\put(6852,402){\makebox(0,0)[lb]{\smash{{\SetFigFont{12}{14.4}{\rmdefault}{\mddefault}{\updefault}$x_2$}}}}
\put(7752,402){\makebox(0,0)[lb]{\smash{{\SetFigFont{12}{14.4}{\rmdefault}{\mddefault}{\updefault}$x_3$}}}}
\put(7302,1302){\makebox(0,0)[lb]{\smash{{\SetFigFont{12}{14.4}{\rmdefault}{\mddefault}{\updefault}$y_2$}}}}
\put(6852,2202){\makebox(0,0)[lb]{\smash{{\SetFigFont{12}{14.4}{\rmdefault}{\mddefault}{\updefault}$z_1$}}}}
\put(5952,402){\makebox(0,0)[lb]{\smash{{\SetFigFont{12}{14.4}{\rmdefault}{\mddefault}{\updefault}$x_1$}}}}
\put(6447,87){\makebox(0,0)[lb]{\smash{{\SetFigFont{14}{16.8}{\rmdefault}{\mddefault}{\updefault}fig.
6c}}}} \path(7077,1617)(7122,1527)
\whiten\path(7041.502,1620.915)(7122.000,1527.000)(7095.167,1647.748)(7041.502,1620.915)
\path(6897,1977)(6942,1887)
\whiten\path(6861.502,1980.915)(6942.000,1887.000)(6915.167,2007.748)(6861.502,1980.915)
\path(6627,717)(6672,627)
\whiten\path(6591.502,720.915)(6672.000,627.000)(6645.167,747.748)(6591.502,720.915)
\path(6447,1077)(6492,987)
\whiten\path(6411.502,1080.915)(6492.000,987.000)(6465.167,1107.748)(6411.502,1080.915)
\path(7527,717)(7572,627)
\whiten\path(7491.502,720.915)(7572.000,627.000)(7545.167,747.748)(7491.502,720.915)
\path(7347,1077)(7392,987)
\whiten\path(7311.502,1080.915)(7392.000,987.000)(7365.167,1107.748)(7311.502,1080.915)
\end{picture}
} \\[8pt]

\noindent \setlength{\unitlength}{0.0008in}
\begingroup\makeatletter\ifx\SetFigFont\undefined%
\gdef\SetFigFont#1#2#3#4#5{%
  \reset@font\fontsize{#1}{#2pt}%
  \fontfamily{#3}\fontseries{#4}\fontshape{#5}%
  \selectfont}%
\fi\endgroup%
{\renewcommand{\dashlinestretch}{30}
\begin{picture}(7588,2817)(0,-10)
\path(1272,2247)(1497,2697) \path(1272,2247)(1497,2697)
\path(1497,2697)(1452,2607) \path(1497,2697)(1452,2607)
\whiten\path(1478.833,2727.748)(1452.000,2607.000)(1532.498,2700.915)(1478.833,2727.748)
\path(1722,1347)(1947,1797) \path(1722,1347)(1947,1797)
\path(1947,1797)(1902,1707) \path(1947,1797)(1902,1707)
\whiten\path(1928.833,1827.748)(1902.000,1707.000)(1982.498,1800.915)(1928.833,1827.748)
\path(2172,447)(2397,897) \path(2172,447)(2397,897)
\path(2397,897)(2352,807) \path(2397,897)(2352,807)
\whiten\path(2378.833,927.748)(2352.000,807.000)(2432.498,900.915)(2378.833,927.748)
\path(822,1347)(1092,1887)
\path(1065.167,1766.252)(1092.000,1887.000)(1011.502,1793.085)
\path(822,1347)(1272,2247) \path(1137,1977)(1092,1887)
\whiten\path(1118.833,2007.748)(1092.000,1887.000)(1172.498,1980.915)(1118.833,2007.748)
\path(957,1617)(912,1527)
\whiten\path(938.833,1647.748)(912.000,1527.000)(992.498,1620.915)(938.833,1647.748)
\path(372,447)(642,987)
\path(615.167,866.252)(642.000,987.000)(561.502,893.085)
\path(372,447)(822,1347) \path(687,1077)(642,987)
\whiten\path(668.833,1107.748)(642.000,987.000)(722.498,1080.915)(668.833,1107.748)
\path(507,717)(462,627)
\whiten\path(488.833,747.748)(462.000,627.000)(542.498,720.915)(488.833,747.748)
\path(1272,447)(1542,987)
\path(1515.167,866.252)(1542.000,987.000)(1461.502,893.085)
\path(1272,447)(1722,1347) \path(1587,1077)(1542,987)
\whiten\path(1568.833,1107.748)(1542.000,987.000)(1622.498,1080.915)(1568.833,1107.748)
\path(1407,717)(1362,627)
\whiten\path(1388.833,747.748)(1362.000,627.000)(1442.498,720.915)(1388.833,747.748)
\put(4242,1347){\blacken\ellipse{46}{46}}
\put(4242,1347){\ellipse{46}{46}}
\put(3342,1347){\blacken\ellipse{46}{46}}
\put(3342,1347){\ellipse{46}{46}}
\put(3792,2247){\blacken\ellipse{46}{46}}
\put(3792,2247){\ellipse{46}{46}} \path(4692,447)(4917,897)
\path(4692,447)(4917,897) \path(4917,897)(4872,807)
\path(4917,897)(4872,807)
\whiten\path(4898.833,927.748)(4872.000,807.000)(4952.498,900.915)(4898.833,927.748)
\path(4242,1347)(4467,1797) \path(4242,1347)(4467,1797)
\path(4467,1797)(4422,1707) \path(4467,1797)(4422,1707)
\whiten\path(4448.833,1827.748)(4422.000,1707.000)(4502.498,1800.915)(4448.833,1827.748)
\path(3792,2247)(4017,2697) \path(3792,2247)(4017,2697)
\path(4017,2697)(3972,2607) \path(4017,2697)(3972,2607)
\whiten\path(3998.833,2727.748)(3972.000,2607.000)(4052.498,2700.915)(3998.833,2727.748)
\path(4242,1347)(4512,807)
\path(4431.502,900.915)(4512.000,807.000)(4485.167,927.748)
\path(4242,1347)(4692,447) \path(3792,2247)(4062,1707)
\path(3981.502,1800.915)(4062.000,1707.000)(4035.167,1827.748)
\path(3792,2247)(4242,1347) \path(3342,1347)(3612,807)
\path(3531.502,900.915)(3612.000,807.000)(3585.167,927.748)
\path(3342,1347)(3792,447) \path(4242,1347)(3792,447)
\path(4242,1347)(3792,447) \path(4242,1347)(3972,807)
\path(4242,1347)(3972,807)
\whiten\path(3998.833,927.748)(3972.000,807.000)(4052.498,900.915)(4041.765,946.531)(3998.833,927.748)
\path(3792,2247)(3342,1347) \path(3792,2247)(3342,1347)
\path(3792,2247)(3522,1707) \path(3792,2247)(3522,1707)
\whiten\path(3548.833,1827.748)(3522.000,1707.000)(3602.498,1800.915)(3591.765,1846.531)(3548.833,1827.748)
\path(3342,1347)(2892,447) \path(3342,1347)(2892,447)
\path(3342,1347)(3072,807) \path(3342,1347)(3072,807)
\whiten\path(3098.833,927.748)(3072.000,807.000)(3152.498,900.915)(3141.765,946.531)(3098.833,927.748)
\put(3477,87){\makebox(0,0)[lb]{\smash{{\SetFigFont{14}{16.8}{\rmdefault}{\mddefault}{\updefault}fig.
6e}}}}
\put(3882,402){\makebox(0,0)[lb]{\smash{{\SetFigFont{12}{14.4}{\rmdefault}{\mddefault}{\updefault}$x_2$}}}}
\put(4782,402){\makebox(0,0)[lb]{\smash{{\SetFigFont{12}{14.4}{\rmdefault}{\mddefault}{\updefault}$x_3$}}}}
\put(3432,1347){\makebox(0,0)[lb]{\smash{{\SetFigFont{12}{14.4}{\rmdefault}{\mddefault}{\updefault}$y_1$}}}}
\put(4332,1347){\makebox(0,0)[lb]{\smash{{\SetFigFont{12}{14.4}{\rmdefault}{\mddefault}{\updefault}$y_2$}}}}
\put(3882,2202){\makebox(0,0)[lb]{\smash{{\SetFigFont{12}{14.4}{\rmdefault}{\mddefault}{\updefault}$z_1$}}}}
\put(2982,402){\makebox(0,0)[lb]{\smash{{\SetFigFont{12}{14.4}{\rmdefault}{\mddefault}{\updefault}$x_1$}}}}
\put(6852,1842){\blacken\ellipse{46}{46}}
\put(6852,1842){\ellipse{46}{46}}
\put(5952,1842){\blacken\ellipse{46}{46}}
\put(5952,1842){\ellipse{46}{46}}
\put(6402,942){\blacken\ellipse{46}{46}}
\put(6402,942){\ellipse{46}{46}} \path(7302,2742)(7527,2292)
\path(7302,2742)(7527,2292) \path(7527,2292)(7482,2382)
\path(7527,2292)(7482,2382)
\whiten\path(7562.498,2288.085)(7482.000,2382.000)(7508.833,2261.252)(7562.498,2288.085)
\path(6852,1842)(7077,1392) \path(6852,1842)(7077,1392)
\path(7077,1392)(7032,1482) \path(7077,1392)(7032,1482)
\whiten\path(7112.498,1388.085)(7032.000,1482.000)(7058.833,1361.252)(7112.498,1388.085)
\path(6402,942)(6627,492) \path(6402,942)(6627,492)
\path(6627,492)(6582,582) \path(6627,492)(6582,582)
\whiten\path(6662.498,488.085)(6582.000,582.000)(6608.833,461.252)(6662.498,488.085)
\path(6852,1842)(7122,2382)
\path(7095.167,2261.252)(7122.000,2382.000)(7041.502,2288.085)
\path(6852,1842)(7302,2742) \path(6402,942)(6672,1482)
\path(6645.167,1361.252)(6672.000,1482.000)(6591.502,1388.085)
\path(6402,942)(6852,1842) \path(5952,1842)(6222,2382)
\path(6195.167,2261.252)(6222.000,2382.000)(6141.502,2288.085)
\path(5952,1842)(6402,2742) \path(6852,1842)(6402,2742)
\path(6852,1842)(6402,2742) \path(6852,1842)(6582,2382)
\path(6852,1842)(6582,2382)
\whiten\path(6662.498,2288.085)(6582.000,2382.000)(6608.833,2261.252)(6651.765,2242.469)(6662.498,2288.085)
\path(6402,942)(5952,1842) \path(6402,942)(5952,1842)
\path(6402,942)(6132,1482) \path(6402,942)(6132,1482)
\whiten\path(6212.498,1388.085)(6132.000,1482.000)(6158.833,1361.252)(6201.765,1342.469)(6212.498,1388.085)
\path(5952,1842)(5502,2742) \path(5952,1842)(5502,2742)
\path(5952,1842)(5682,2382) \path(5952,1842)(5682,2382)
\whiten\path(5762.498,2288.085)(5682.000,2382.000)(5708.833,2261.252)(5751.765,2242.469)(5762.498,2288.085)
\put(6087,87){\makebox(0,0)[lb]{\smash{{\SetFigFont{14}{16.8}{\rmdefault}{\mddefault}{\updefault}fig.
6f}}}}
\put(6492,2697){\makebox(0,0)[lb]{\smash{{\SetFigFont{12}{14.4}{\rmdefault}{\mddefault}{\updefault}$x_2$}}}}
\put(7392,2697){\makebox(0,0)[lb]{\smash{{\SetFigFont{12}{14.4}{\rmdefault}{\mddefault}{\updefault}$x_3$}}}}
\put(6042,1797){\makebox(0,0)[lb]{\smash{{\SetFigFont{12}{14.4}{\rmdefault}{\mddefault}{\updefault}$y_1$}}}}
\put(6942,1797){\makebox(0,0)[lb]{\smash{{\SetFigFont{12}{14.4}{\rmdefault}{\mddefault}{\updefault}$y_2$}}}}
\put(6492,897){\makebox(0,0)[lb]{\smash{{\SetFigFont{12}{14.4}{\rmdefault}{\mddefault}{\updefault}$z_1$}}}}
\put(5592,2697){\makebox(0,0)[lb]{\smash{{\SetFigFont{12}{14.4}{\rmdefault}{\mddefault}{\updefault}$x_1$}}}}
\put(822,1347){\blacken\ellipse{46}{46}}
\put(822,1347){\ellipse{46}{46}}
\put(1722,1347){\blacken\ellipse{46}{46}}
\put(1722,1347){\ellipse{46}{46}}
\put(1272,2247){\blacken\ellipse{46}{46}}
\put(1272,2247){\ellipse{46}{46}} \path(822,1347)(1092,807)
\path(1011.502,900.915)(1092.000,807.000)(1065.167,927.748)
\path(1722,1347)(1992,807)
\path(1911.502,900.915)(1992.000,807.000)(1965.167,927.748)
\path(1272,2247)(1542,1707)
\path(1461.502,1800.915)(1542.000,1707.000)(1515.167,1827.748)
\path(1272,447)(822,1347) \path(2172,447)(1722,1347)
\path(1722,1347)(1272,2247) \path(2397,1527)(2892,1527)
\blacken\path(2772.000,1497.000)(2892.000,1527.000)(2772.000,1557.000)(2808.000,1527.000)(2772.000,1497.000)
\path(5007,1527)(5502,1527)
\blacken\path(5382.000,1497.000)(5502.000,1527.000)(5382.000,1557.000)(5418.000,1527.000)(5382.000,1497.000)
\path(12,1527)(507,1527)
\blacken\path(387.000,1497.000)(507.000,1527.000)(387.000,1557.000)(423.000,1527.000)(387.000,1497.000)
\put(912,1302){\makebox(0,0)[lb]{\smash{{\SetFigFont{12}{14.4}{\rmdefault}{\mddefault}{\updefault}$y_1$}}}}
\put(1362,402){\makebox(0,0)[lb]{\smash{{\SetFigFont{12}{14.4}{\rmdefault}{\mddefault}{\updefault}$x_2$}}}}
\put(2262,402){\makebox(0,0)[lb]{\smash{{\SetFigFont{12}{14.4}{\rmdefault}{\mddefault}{\updefault}$x_3$}}}}
\put(1812,1302){\makebox(0,0)[lb]{\smash{{\SetFigFont{12}{14.4}{\rmdefault}{\mddefault}{\updefault}$y_2$}}}}
\put(1362,2202){\makebox(0,0)[lb]{\smash{{\SetFigFont{12}{14.4}{\rmdefault}{\mddefault}{\updefault}$z_1$}}}}
\put(462,402){\makebox(0,0)[lb]{\smash{{\SetFigFont{12}{14.4}{\rmdefault}{\mddefault}{\updefault}$x_1$}}}}
\put(957,87){\makebox(0,0)[lb]{\smash{{\SetFigFont{14}{16.8}{\rmdefault}{\mddefault}{\updefault}fig.
6d}}}}
\end{picture}
} \\[8pt]

To obtain a graphical representation for the integral in~\eqref{orthog} we should join the diagrams
shown in the figs.~6a for $\psi_{\gamma}(x)$ and 6f for
$\overline{\psi_{\gamma}(x)}$ in the points $x_1,\ldots,x_N$ and
integrating over them. The diagram obtained like this
in the case $N=4$ is pictured in fig.~8a. Further we shall
simplify it using the following equalities.

\begin{lemma} \label{lemma2}
  If $\gamma'_j\ne\gamma_k$ the equalities represented in the figures~7a and 7b are valid.
\end{lemma}

\hspace{\fill} \setlength{\unitlength}{0.00087489in}
\begingroup\makeatletter\ifx\SetFigFont\undefined%
\gdef\SetFigFont#1#2#3#4#5{%
  \reset@font\fontsize{#1}{#2pt}%
  \fontfamily{#3}\fontseries{#4}\fontshape{#5}%
  \selectfont}%
\fi\endgroup%
{\renewcommand{\dashlinestretch}{30}
\begin{picture}(4366,2565)(0,-10)
\put(3210,1122){\makebox(0,0)[lb]{\smash{{\SetFigFont{10}{14.4}{\rmdefault}{\mddefault}{\updefault}$\gamma_k$}}}}
\put(3750,1572){\makebox(0,0)[lb]{\smash{{\SetFigFont{10}{12.0}{\rmdefault}{\mddefault}{\updefault}\makebox[0.8\width][s]{$(\gamma'_j-\gamma_k)$}}}}}
\path(3660,672)(3390,1212) \path(3660,672)(3390,1212)
\whiten\path(3470.498,1118.085)(3390.000,1212.000)(3416.833,1091.252)(3470.498,1118.085)
\path(3660,672)(3660,2472) \path(3660,672)(3660,2472)
\path(3660,672)(3660,1707) \path(3660,672)(3660,1707)
\whiten\path(3690.000,1587.000)(3660.000,1707.000)(3630.000,1587.000)(3660.000,1623.000)(3690.000,1587.000)
\put(3750,2427){\makebox(0,0)[lb]{\smash{{\SetFigFont{12}{14.4}{\rmdefault}{\mddefault}{\updefault}$y$}}}}
\put(3750,627){\makebox(0,0)[lb]{\smash{{\SetFigFont{12}{14.4}{\rmdefault}{\mddefault}{\updefault}$y'$}}}}
\put(15,2022){\makebox(0,0)[lb]{\smash{{\SetFigFont{10}{14.4}{\rmdefault}{\mddefault}{\updefault}$\gamma_k$}}}}
\put(555,1527){\makebox(0,0)[lb]{\smash{{\SetFigFont{12}{14.4}{\rmdefault}{\mddefault}{\updefault}$x$}}}}
\path(465,1572)(195,2112) \path(465,1572)(195,2112)
\whiten\path(275.498,2018.085)(195.000,2112.000)(221.833,1991.252)(275.498,2018.085)
\path(465,1572)(735,2112)
\path(708.167,1991.252)(735.000,2112.000)(654.502,2018.085)
\path(465,1572)(915,2472) \path(465,1572)(735,2112)
\path(708.167,1991.252)(735.000,2112.000)(654.502,2018.085)
\path(465,1572)(915,2472) \path(915,672)(465,1572)
\path(915,672)(465,1572) \path(915,672)(645,1212)
\path(915,672)(645,1212)
\whiten\path(725.498,1118.085)(645.000,1212.000)(671.833,1091.252)(714.765,1072.469)(725.498,1118.085)
\put(465,1122){\makebox(0,0)[lb]{\smash{{\SetFigFont{10}{14.4}{\rmdefault}{\mddefault}{\updefault}$\gamma'_j$}}}}
\put(1005,2427){\makebox(0,0)[lb]{\smash{{\SetFigFont{12}{14.4}{\rmdefault}{\mddefault}{\updefault}$y$}}}}
\put(1005,627){\makebox(0,0)[lb]{\smash{{\SetFigFont{12}{14.4}{\rmdefault}{\mddefault}{\updefault}$y'$}}}}
\put(465,1572){\blacken\ellipse{46}{46}}
\put(465,1572){\ellipse{46}{46}}
\put(1320,1527){\makebox(0,0)[lb]{\smash{{\SetFigFont{14}{16.8}{\rmdefault}{\mddefault}{\updefault}=}}}}
\put(1590,1527){\makebox(0,0)[lb]{\smash{{\SetFigFont{14}{16.8}{\rmdefault}{\mddefault}{\updefault}$\hbar^{\frac{\gamma'_j-\gamma_k}{i\hbar}}\Gamma\Bigl(\dfrac{\gamma'_j-\gamma_k}{i\hbar}\Bigr)$}}}}
\put(1995,87){\makebox(0,0)[lb]{\smash{{\SetFigFont{14}{16.8}{\rmdefault}{\mddefault}{\updefault}fig.
7a}}}}
\end{picture}
} \hspace{\fill} \\

\hspace{\fill} \setlength{\unitlength}{0.00087489in}
\begingroup\makeatletter\ifx\SetFigFont\undefined%
\gdef\SetFigFont#1#2#3#4#5{%
  \reset@font\fontsize{#1}{#2pt}%
  \fontfamily{#3}\fontseries{#4}\fontshape{#5}%
  \selectfont}%
\fi\endgroup%
{\renewcommand{\dashlinestretch}{30}
\begin{picture}(4051,2430)(0,-10)
\put(1095,1392){\makebox(0,0)[lb]{\smash{{\SetFigFont{14}{16.8}{\rmdefault}{\mddefault}{\updefault}=}}}}
\put(1365,1392){\makebox(0,0)[lb]{\smash{{\SetFigFont{14}{16.8}{\rmdefault}{\mddefault}{\updefault}$\hbar^{\frac{\gamma_k-\gamma'_j}{i\hbar}}\Gamma\Bigl(\dfrac{\gamma_k-\gamma'_j}{i\hbar}\Bigr)$}}}}
\path(3345,537)(3345,2337) \path(3345,537)(3345,2337)
\path(3345,537)(3345,1572) \path(3345,537)(3345,1572)
\whiten\path(3375.000,1452.000)(3345.000,1572.000)(3315.000,1452.000)(3345.000,1488.000)(3375.000,1452.000)
\put(3660,1842){\makebox(0,0)[lb]{\smash{{\SetFigFont{10}{14.4}{\rmdefault}{\mddefault}{\updefault}$\gamma'_j$}}}}
\path(3345,2337)(3570,1887) \path(3345,2337)(3570,1887)
\path(465,1437)(15,2337) \path(465,1437)(15,2337)
\path(465,1437)(195,1977) \path(465,1437)(195,1977)
\whiten\path(275.498,1883.085)(195.000,1977.000)(221.833,1856.252)(264.765,1837.469)(275.498,1883.085)
\put(15,1887){\makebox(0,0)[lb]{\smash{{\SetFigFont{10}{14.4}{\rmdefault}{\mddefault}{\updefault}$\gamma_k$}}}}
\path(15,537)(285,1077)
\path(258.167,956.252)(285.000,1077.000)(204.502,983.085)
\path(15,537)(465,1437)
\put(465,987){\makebox(0,0)[lb]{\smash{{\SetFigFont{10}{14.4}{\rmdefault}{\mddefault}{\updefault}$\gamma'_j$}}}}
\put(105,2292){\makebox(0,0)[lb]{\smash{{\SetFigFont{12}{14.4}{\rmdefault}{\mddefault}{\updefault}$y$}}}}
\put(105,492){\makebox(0,0)[lb]{\smash{{\SetFigFont{12}{14.4}{\rmdefault}{\mddefault}{\updefault}$y'$}}}}
\put(555,1392){\makebox(0,0)[lb]{\smash{{\SetFigFont{12}{14.4}{\rmdefault}{\mddefault}{\updefault}$x$}}}}
\put(465,1437){\blacken\ellipse{46}{46}}
\put(465,1437){\ellipse{46}{46}} \path(465,1437)(690,987)
\path(465,1437)(690,987) \path(690,987)(645,1077)
\path(690,987)(645,1077)
\whiten\path(725.498,983.085)(645.000,1077.000)(671.833,956.252)(725.498,983.085)
\put(1725,87){\makebox(0,0)[lb]{\smash{{\SetFigFont{14}{16.8}{\rmdefault}{\mddefault}{\updefault}fig.
7b}}}}
\put(3435,2292){\makebox(0,0)[lb]{\smash{{\SetFigFont{12}{14.4}{\rmdefault}{\mddefault}{\updefault}$y$}}}}
\put(3435,1437){\makebox(0,0)[lb]{\smash{{\SetFigFont{10}{12.0}{\rmdefault}{\mddefault}{\updefault}\makebox[0.8\width][s]{$(\gamma_k-\gamma'_j)$}}}}}
\put(3435,492){\makebox(0,0)[lb]{\smash{{\SetFigFont{12}{14.4}{\rmdefault}{\mddefault}{\updefault}$y'$}}}}
\path(3570,1887)(3525,1977) \path(3570,1887)(3525,1977)
\whiten\path(3605.498,1883.085)(3525.000,1977.000)(3551.833,1856.252)(3605.498,1883.085)
\end{picture}
} \hspace{\fill} \\

\noindent{\bfseries Proof.} The fig.~7a and 7b mean the following
equalities, which are consequence of the integral representation
of the Gamma-function
\begin{multline*}
 \int\limits_{-\infty}^{+\infty} dx\, Z_{\gamma_k}(x)I(x,y)J_{\gamma'_j}(y',x)=
 \int\limits_{-\infty}^{+\infty} dx\, \exp\Bigl\{\frac{i}{\hbar}[(\gamma_k-\gamma'_j)x+\gamma'_j y']
   -\frac{1}{\hbar}(e^{x-y}+e^{x-y'})\Bigr\}=\\
 =e^{\frac{i}{\hbar}\gamma'_j y'}
  \left(\frac{\hbar}{e^{-y}+e^{-y'}}\right)^{\frac{i}{\hbar}(\gamma_k-\gamma'_j)}
  \Gamma\Bigl(\frac{\gamma'_j-\gamma_k}{i\hbar}\Bigr)=
  \hbar^{\frac{\gamma'_j-\gamma_k}{i\hbar}}\Gamma\Bigl(\frac{\gamma'_j-\gamma_k}{i\hbar}\Bigr)
  Z_{\gamma_k}(y') Y_{\gamma'_j-\gamma_k}(y',y),
\end{multline*}
\begin{multline*}
 \int\limits_{-\infty}^{+\infty} dx\, J_{\gamma_k}(x,y)I(y',x)Z_{\gamma'_j}^{-1}(x)=
 \int\limits_{-\infty}^{+\infty} dx\, \exp\Bigl\{\frac{i}{\hbar}[(\gamma_k-\gamma'_j)x-\gamma_k y]
   -\frac{1}{\hbar}(e^{y-x}+e^{y'-x})\Bigr\}=\\
 =e^{-\frac{i}{\hbar}\gamma_k y}
  \left(\frac{\hbar}{e^{y}+e^{y'}}\right)^{\frac{i}{\hbar}(\gamma'_j-\gamma_k)}
  \Gamma\Bigl(\frac{\gamma_k-\gamma'_j}{i\hbar}\Bigr)=
  \hbar^{\frac{\gamma_k-\gamma'_j}{i\hbar}}\Gamma\Bigl(\frac{\gamma_k-\gamma'_j}{i\hbar}\Bigr)
  Y_{\gamma_k-\gamma'_j}(y',y) Z_{\gamma'_j}(y).
\end{multline*}

The poles of Gamma-function are non-positive integers. Since the variables $\gamma'_j$ and $\gamma_k$ are not equal to each other and have only real values the right-hand sides of the equalities encoded in the figures~7a and 7b are defined correctly.
\qed \\

The calculation of the integral in the left hand side of~\eqref{orthog} can be reduced to a calculation of this integral in the domain
$\gamma'_j\ne\gamma_k$, $j+k\ne N+1$. In order to do it we need a smooth partition of unity which is symmetric over $\gamma$ with a component vanishing in some neihgbourhood of the plane $\gamma'_j=\gamma_k$, $j+k\ne N+1$. \\

Since the variables $\gamma_k$ are pairwise different one can choose a positive number $\epsilon>0$ such that $6\epsilon<\min_{k\ne l}|\gamma_k-\gamma_l|$. Let $\eta_\epsilon(x)$ be a smooth function of one variable such that $0\le\eta_\epsilon(x)\le1$ for all $x\in\mathbb R$, $\eta_\epsilon(x)=1$ for $|x|<\epsilon$, $\eta_\epsilon(x)=0$ for $|x|>3\epsilon$. Let us set
\begin{align}
 \tilde\phi(\gamma';\gamma)&=\prod_{\substack{j,k=1 \\ j+k\ne N+1}}^N \Big(1-\eta_\epsilon(\gamma'_j-\gamma_k)\Big), \notag \\
 \phi(\gamma';\gamma)&=\frac{\tilde\phi(\gamma';\gamma)}{\sum\limits_{\sigma\in S_N}\tilde\phi(\gamma';\sigma\gamma)},      \label{phi}
\end{align}
where $S_N$ is the permutation group and $\sigma\gamma=(\gamma_{\sigma(1)},\ldots,\gamma_{\sigma(N)})$. 
The formula~\eqref{phi} defines corectly a smooth function. To prove this it is sufficient to show that the denominator in the right hand side of this formula never vanishes. Since the functions $\tilde\phi(\gamma';\sigma\gamma)$ are non-negative we have to prove that for all $\gamma'\in\mathbb R^N$ there exists a permutation $\sigma\in S_N$ such that $\tilde\phi(\gamma';\sigma\gamma)\ne0$. Now fix $\gamma'\in\mathbb R^N$ and define this permutation $\sigma\in S_N$ in the following way: let us consider a subset of subscript pairs ${\mathfrak S}(\gamma')=\{(j,k)\mid|\gamma'_j-\gamma_k|<3\epsilon\}$. It has the property following: if $(j,k)\in{\mathfrak S}(\gamma')$ and $(j,l)\in{\mathfrak S}(\gamma')$ then $k=l$. Indeed, if $k\ne l$ we have a chain of inequalities $6\epsilon<|\gamma_k-\gamma_l|\le|\gamma_k-\gamma'_j|+|\gamma'_j-\gamma_l|<3\epsilon+3\epsilon=6\epsilon$, which leads to contradiction. This property means that there exist a permutation $\tilde\sigma\in S_N$ such that $j=\tilde\sigma(k)$ for all $(j,k)\in{\mathfrak S}(\gamma')$. Now let us show that $\tilde\phi(\gamma';\sigma\gamma)\ne0$ for $\sigma=\tilde\sigma^{-1}\alpha$, where $\alpha\in S_N$ is the longest permutation: $\alpha(j)=N+1-j$. Noting that $k=\sigma(N+1-j)$ for all $(j,k)\in{\mathfrak S}(\gamma')$ we see that $|\gamma'_j-\gamma_{\sigma(N+1-m)}|\ge3\epsilon$ for all $j\ne m$.
Hence
\begin{align*}
 \tilde\phi(\gamma';\sigma\gamma)=\prod_{j=1}^N\prod_{\substack{m=1 \\ m\ne j}}^N \Big(1-\eta_\epsilon(\gamma'_j-\gamma_{\sigma(N+1-m)})\Big)=1\ne0.
\end{align*}

The function~\eqref{phi} has the following properties:
\begin{enumerate}
 \item\label{prop1} $\sum\limits_{\sigma\in S_N}\phi(\gamma';\sigma\gamma)=1$ (a partition of unity),
 \item\label{prop2} if $|\gamma'_j-\gamma_k|<\epsilon$ for some $j,k$ such that $j+k\ne N+1$ then $\phi(\gamma';\gamma)=0$,
 \item\label{prop3} if $|\gamma'_j-\gamma_{N+1-j}|<\epsilon$ for all $j=1,\ldots,N$ then $\phi(\gamma';\gamma)=1$.
\end{enumerate}

The property~\ref{prop1} is obvious. The property~\ref{prop2} follows from the fact that if $|\gamma'_j-\gamma_k|<\epsilon$ for some $j,k$ such that $j+k\ne N+1$ then $\tilde\phi(\gamma';\gamma)=0$. The property~\ref{prop2} follows from the same fact. Indeed, fix $\gamma'\in\mathbb R^N$ and suppose that $|\gamma'_j-\gamma_{N+1-j}|<\epsilon$ for all $j=1,\ldots,N$. If $\sigma\ne1$ then there exist such $j$ that $\sigma^{-1}(N+1-j)\ne N+1-j$. Denoting $k=\sigma^{-1}(N+1-j)$ one derives $|\gamma_j-\gamma_{\sigma(k)}|<0$ and hence $\tilde\phi(\gamma';\sigma\gamma)=0$. In the other hand we have proved that there exists $\sigma\in S_N$ such that $\tilde\phi(\gamma';\sigma\gamma)\ne0$, that is $\tilde\phi(\gamma';\gamma)\ne0$. Hence
\begin{align*}
 \phi(\gamma';\gamma)=\frac{\tilde\phi(\gamma';\gamma)}{\tilde\phi(\gamma';\gamma)}=1.
\end{align*}

Due to the property 1 and the symmetry of the sought integral the last one can be rewritten in the form
\begin{align*}
 \int\limits_{\mathbb R^N}dx\overline{\psi_\gamma(x)}\psi_{\gamma'}(x)=\sum\limits_{\sigma\in S_N}{\cal I}(\gamma';\sigma\gamma),
\end{align*}
where
\begin{align*}
 {\cal I}(\gamma';\gamma)=\phi(\gamma';\gamma)\int\limits_{\mathbb R^N}dx\overline{\psi_\gamma(x)}\psi_{\gamma'}(x).
\end{align*}
Therefore, since the Sklyanin measure $\mu(\gamma)$ is symmetric over $\gamma$ it is sufficient to derive the equality
\begin{align}
 {\cal I}(\gamma';\gamma)=\mu^{-1}(\gamma)\frac1{N!}\prod_{j=1}^N\delta(\gamma'_j-\gamma_{N+1-j}). \label{I_delta}
\end{align}
The property~\ref{prop2} means that calculating ${\cal I}(\gamma';\gamma)$ one can use the lemma~\ref{lemma2} for $j+k\ne N+1$, because the poles of Gamma-functions will be nulled by function $\phi(\gamma';\gamma)$. The property~\ref{prop3} will be needed further. \\

The expression ${\cal I}(\gamma';\gamma)$ is the integral represented graphically in the fig.~8a multiplied by $\phi(\gamma';\gamma)$. It can be calculated by induction. We integrate step by step over the
boundary points connected with $\gamma_1$, $\gamma'_1$. First, we
integrate over the extremely left point and the extremely right point, i.e.
over $x_1$ and $x_N$, using the fig.~7a and 7b respectively, and
we obtain the fig.~8b. Like in the fig.~4, the vertical line arising in
the left side is moved to right, where it is annihilated by the
line arising from the right side. This process exchanges
$\gamma_1$ with $\gamma'_1$. After the integration the factor
\begin{equation} \label{fac_1}
 \Gamma\Bigl(\frac{\gamma'_1-\gamma_1}{i\hbar}\Bigr)
 \Gamma\Bigl(\frac{\gamma_1-\gamma'_1}{i\hbar}\Bigr)
\end{equation}
arises, and thus one obtains the fig.~8c, where $\gamma_1$,
$\gamma'_1$ begin to be connected with another boundary points
(boundary in sense of fig.~8a), 
Now we integrate over these points. In $k$-th integration
($k=2,\ldots,N-1$) $\gamma_1$ exchanges with $\gamma'_k$ and
$\gamma'_1$ exchanges with $\gamma_k$ and one has the factor
\begin{equation} \label{fac_k}
 \Gamma\Bigl(\frac{\gamma'_1-\gamma_k}{i\hbar}\Bigr)
 \Gamma\Bigl(\frac{\gamma_k-\gamma'_1}{i\hbar}\Bigr)
 \Gamma\Bigl(\frac{\gamma'_k-\gamma_1}{i\hbar}\Bigr)
 \Gamma\Bigl(\frac{\gamma_1-\gamma'_k}{i\hbar}\Bigr).
\end{equation}
The $N$-th integration leads to the factor
\begin{equation} \label{fac_N}
 (2\pi\hbar)\delta(\gamma'_1-\gamma_N)\cdot(2\pi\hbar)\delta(\gamma'_N-\gamma_1).
\end{equation}
After this the variables $\gamma_1$, $\gamma'_1$, $\gamma_N$ and
$\gamma'_N$ disappear completely from diagram -- they have gone
away to the factors~\eqref{fac_1}, \eqref{fac_k} and
\eqref{fac_N}. The remaining diagram (the middle part of fig.~8e)
is exactly the initial diagram, but for $N-2$ which is depend on
$\gamma_2,\ldots,\gamma_{N-1}$, $\gamma'_2,\ldots,\gamma'_{N-1}$.
Thereby we have obtained the recursion formula
\begin{multline*} 
 \phi(\gamma';\gamma)\int dx_1\ldots dx_N\,\overline{\psi_{\gamma_1,\ldots,\gamma_N}(x_1,\ldots,x_N)}
                                 \psi_{\gamma'_1,\ldots,\gamma'_N}(x_1,\ldots,x_N)= \\
 =\phi(\gamma';\gamma)
 \Gamma\Bigl(\frac{\gamma'_1-\gamma_1}{i\hbar}\Bigr)\Gamma\Bigl(\frac{\gamma_1-\gamma'_1}{i\hbar}\Bigr)\times \\
 \times\prod_{k=2}^{N-1}\Gamma\Bigl(\frac{\gamma'_1-\gamma_k}{i\hbar}\Bigr)\Gamma\Bigl(\frac{\gamma_k-\gamma'_1}{i\hbar}\Bigr)
 \Gamma\Bigl(\frac{\gamma'_k-\gamma_1}{i\hbar}\Bigr)\Gamma\Bigl(\frac{\gamma_1-\gamma'_k}{i\hbar}\Bigr)\cdot
 (2\pi\hbar)\delta(\gamma'_1-\gamma_N)\times \\
 \times(2\pi\hbar)\delta(\gamma'_N-\gamma_1)
 \int dx_2\ldots dx_{N-1}\,\overline{\psi_{\gamma_2,\ldots,\gamma_{N-1}}(x_2,\ldots,x_{N-1})}
                                     \psi_{\gamma'_2,\ldots,\gamma'_{N-1}}(x_2,\ldots,x_{N-1}).
\end{multline*}
Continuing the calculation for the integral~\eqref{orthog} in the same
manner we obtain the following result.
\begin{multline} \label{int_psi}
 \phi(\gamma';\gamma)\int dx_1\ldots dx_N\,\overline{\psi_{\gamma_1,\ldots,\gamma_N}(x_1,\ldots,x_N)}
                                 \psi_{\gamma'_1,\ldots,\gamma'_N}(x_1,\ldots,x_N)=\\
 =\phi(\gamma';\gamma)\prod_{\substack{k,j=1\\k+j\le N}}^{N-1}\left[\Gamma\Bigl(\frac{\gamma'_j-\gamma_k}{i\hbar}\Bigr)
 \Gamma\Bigl(\frac{\gamma_k-\gamma'_j}{i\hbar}\Bigr)\right] \prod_{j=1}^N (2\pi\hbar) \delta(\gamma'_j-\gamma_{N+1-j}).
\end{multline}

\noindent\hspace{-3mm} \setlength{\unitlength}{0.0004in} 
\begingroup\makeatletter\ifx\SetFigFont\undefined%
\gdef\SetFigFont#1#2#3#4#5{%
  \reset@font\fontsize{#1}{#2pt}%
  \fontfamily{#3}\fontseries{#4}\fontshape{#5}%
  \selectfont}%
\fi\endgroup%
{\renewcommand{\dashlinestretch}{30}
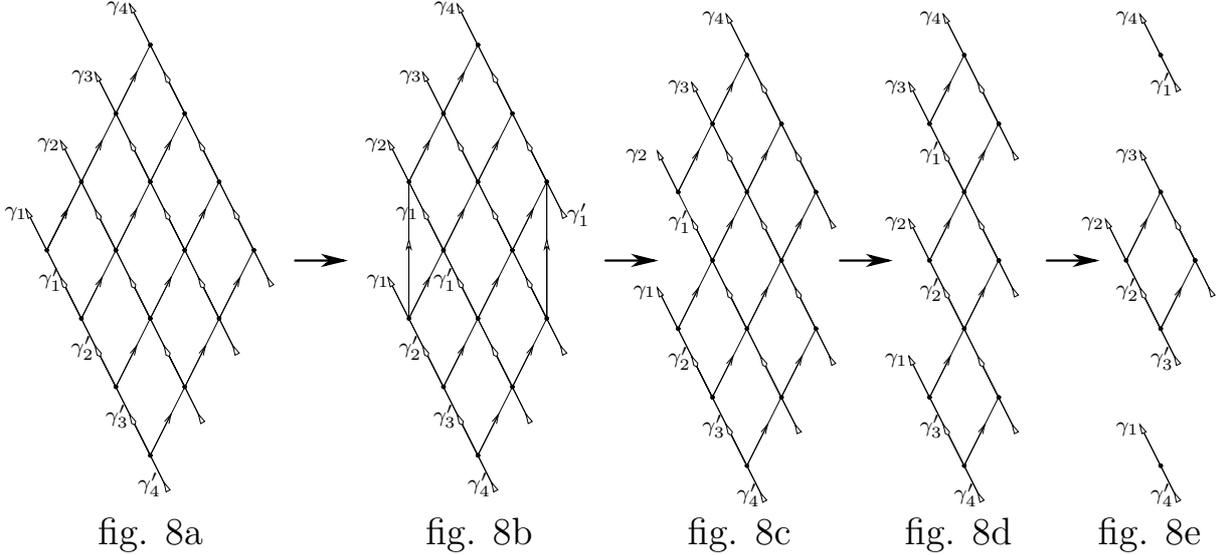
\begin{picture}(8272,7240)(0,-10)
\path(2265,3952)(2535,4492)
\path(2508.167,4371.252)(2535.000,4492.000)(2454.502,4398.085)
\path(2265,3952)(2715,4852) \path(2715,4852)(2265,5752)
\path(2715,4852)(2265,5752) \path(2715,4852)(2445,5392)
\path(2715,4852)(2445,5392)
\whiten\path(2525.498,5298.085)(2445.000,5392.000)(2471.833,5271.252)(2514.765,5252.469)(2525.498,5298.085)
\path(915,4852)(1185,5392)
\path(1158.167,5271.252)(1185.000,5392.000)(1104.502,5298.085)
\path(915,4852)(1365,5752) \path(1365,5752)(1635,6292)
\path(1608.167,6171.252)(1635.000,6292.000)(1554.502,6198.085)
\path(1365,5752)(1815,6652) \path(2265,5752)(1815,6652)
\path(2265,5752)(1815,6652) \path(2265,5752)(1995,6292)
\path(2265,5752)(1995,6292)
\whiten\path(2075.498,6198.085)(1995.000,6292.000)(2021.833,6171.252)(2064.765,6152.469)(2075.498,6198.085)
\put(-85,4402){\makebox(0,0)[lb]{\smash{{\SetFigFont{8}{14.4}{\rmdefault}{\mddefault}{\updefault}$\gamma_1$}}}}
\path(465,3952)(195,4492) \path(465,3952)(195,4492)
\whiten\path(275.498,4398.085)(195.000,4492.000)(221.833,4371.252)(275.498,4398.085)
\path(1815,4852)(2085,5392)
\path(2058.167,5271.252)(2085.000,5392.000)(2004.502,5298.085)
\path(1815,4852)(2265,5752)
\path(1815,4852)(1365,5752) \path(1815,4852)(1365,5752)
\path(1815,4852)(1545,5392) \path(1815,4852)(1545,5392)
\whiten\path(1625.498,5298.085)(1545.000,5392.000)(1571.833,5271.252)(1614.765,5252.469)(1625.498,5298.085)
\path(1365,3952)(1635,4492)
\path(1608.167,4371.252)(1635.000,4492.000)(1554.502,4398.085)
\path(1365,3952)(1815,4852) \path(2265,3952)(1815,4852)
\path(2265,3952)(1815,4852) \path(2265,3952)(1995,4492)
\path(2265,3952)(1995,4492)
\whiten\path(2075.498,4398.085)(1995.000,4492.000)(2021.833,4371.252)(2064.765,4352.469)(2075.498,4398.085)
\put(1815,4852){\blacken\ellipse{46}{46}}
\put(1815,4852){\ellipse{46}{46}}
\path(1365,3952)(915,4852) \path(1365,3952)(915,4852)
\path(1365,3952)(1095,4492) \path(1365,3952)(1095,4492)
\whiten\path(1175.498,4398.085)(1095.000,4492.000)(1121.833,4371.252)(1164.765,4352.469)(1175.498,4398.085)
\put(915,4852){\blacken\ellipse{46}{46}}
\put(915,4852){\ellipse{46}{46}} \path(465,3952)(735,4492)
\path(708.167,4371.252)(735.000,4492.000)(654.502,4398.085)
\path(465,3952)(915,4852) \path(465,3952)(735,4492)
\path(708.167,4371.252)(735.000,4492.000)(654.502,4398.085)
\path(465,3952)(915,4852) \path(915,3052)(1185,3592)
\path(1158.167,3471.252)(1185.000,3592.000)(1104.502,3498.085)
\path(915,3052)(1365,3952) \path(1815,3052)(1365,3952)
\path(1815,3052)(1365,3952) \path(1815,3052)(1545,3592)
\path(1815,3052)(1545,3592)
\whiten\path(1625.498,3498.085)(1545.000,3592.000)(1571.833,3471.252)(1614.765,3452.469)(1625.498,3498.085)
\path(1815,3052)(2085,3592)
\path(2058.167,3471.252)(2085.000,3592.000)(2004.502,3498.085)
\path(1815,3052)(2265,3952) \path(2715,3052)(2985,3592)
\path(2958.167,3471.252)(2985.000,3592.000)(2904.502,3498.085)
\path(2715,3052)(3165,3952) \path(2715,3052)(2265,3952)
\path(2715,3052)(2265,3952) \path(2715,3052)(2445,3592)
\path(2715,3052)(2445,3592)
\whiten\path(2525.498,3498.085)(2445.000,3592.000)(2471.833,3471.252)(2514.765,3452.469)(2525.498,3498.085)
\path(2265,2152)(2535,2692)
\path(2508.167,2571.252)(2535.000,2692.000)(2454.502,2598.085)
\path(2265,2152)(2715,3052) \path(2265,2152)(1815,3052)
\path(2265,2152)(1815,3052) \path(2265,2152)(1995,2692)
\path(2265,2152)(1995,2692)
\whiten\path(2075.498,2598.085)(1995.000,2692.000)(2021.833,2571.252)(2064.765,2552.469)(2075.498,2598.085)
\path(1365,2152)(1635,2692)
\path(1608.167,2571.252)(1635.000,2692.000)(1554.502,2598.085)
\path(1365,2152)(1815,3052) \path(1365,2152)(915,3052)
\path(1365,2152)(915,3052) \path(1365,2152)(1095,2692)
\path(1365,2152)(1095,2692)
\whiten\path(1175.498,2598.085)(1095.000,2692.000)(1121.833,2571.252)(1164.765,2552.469)(1175.498,2598.085)
\path(1815,1252)(2085,1792)
\path(2058.167,1671.252)(2085.000,1792.000)(2004.502,1698.085)
\path(1815,1252)(2265,2152) \path(1815,1252)(1365,2152)
\path(1815,1252)(1365,2152) \path(1815,1252)(1545,1792)
\path(1815,1252)(1545,1792)
\whiten\path(1625.498,1698.085)(1545.000,1792.000)(1571.833,1671.252)(1614.765,1652.469)(1625.498,1698.085)
\path(3165,3952)(2715,4852) \path(3165,3952)(2715,4852)
\path(3165,3952)(2895,4492) \path(3165,3952)(2895,4492)
\whiten\path(2975.498,4398.085)(2895.000,4492.000)(2921.833,4371.252)(2964.765,4352.469)(2975.498,4398.085)
\path(915,3052)(465,3952) \path(915,3052)(465,3952)
\path(915,3052)(645,3592) \path(915,3052)(645,3592)
\whiten\path(725.498,3498.085)(645.000,3592.000)(671.833,3471.252)(714.765,3452.469)(725.498,3498.085)
\path(6540,3952)(6810,4492)
\path(6783.167,4371.252)(6810.000,4492.000)(6729.502,4398.085)
\path(6540,3952)(6990,4852) \path(6990,4852)(6540,5752)
\path(6990,4852)(6540,5752) \path(6990,4852)(6720,5392)
\path(6990,4852)(6720,5392)
\whiten\path(6800.498,5298.085)(6720.000,5392.000)(6746.833,5271.252)(6789.765,5252.469)(6800.498,5298.085)
\path(5190,4852)(5460,5392)
\path(5433.167,5271.252)(5460.000,5392.000)(5379.502,5298.085)
\path(5190,4852)(5640,5752) \path(5640,5752)(5910,6292)
\path(5883.167,6171.252)(5910.000,6292.000)(5829.502,6198.085)
\path(5640,5752)(6090,6652) \path(6540,5752)(6090,6652)
\path(6540,5752)(6090,6652) \path(6540,5752)(6270,6292)
\path(6540,5752)(6270,6292)
\whiten\path(6350.498,6198.085)(6270.000,6292.000)(6296.833,6171.252)(6339.765,6152.469)(6350.498,6198.085)
\path(6090,4852)(6360,5392)
\path(6333.167,5271.252)(6360.000,5392.000)(6279.502,5298.085)
\path(6090,4852)(6540,5752)
\path(6090,4852)(5640,5752) \path(6090,4852)(5640,5752)
\path(6090,4852)(5820,5392) \path(6090,4852)(5820,5392)
\whiten\path(5900.498,5298.085)(5820.000,5392.000)(5846.833,5271.252)(5889.765,5252.469)(5900.498,5298.085)
\path(5190,3052)(5460,3592)
\path(5433.167,3471.252)(5460.000,3592.000)(5379.502,3498.085)
\path(5190,3052)(5640,3952) \path(6090,3052)(5640,3952)
\path(6090,3052)(5640,3952) \path(6090,3052)(5820,3592)
\path(6090,3052)(5820,3592)
\whiten\path(5900.498,3498.085)(5820.000,3592.000)(5846.833,3471.252)(5889.765,3452.469)(5900.498,3498.085)
\path(6090,3052)(6360,3592)
\path(6333.167,3471.252)(6360.000,3592.000)(6279.502,3498.085)
\path(6090,3052)(6540,3952) \path(6990,3052)(6540,3952)
\path(6990,3052)(6540,3952) \path(6990,3052)(6720,3592)
\path(6990,3052)(6720,3592)
\whiten\path(6800.498,3498.085)(6720.000,3592.000)(6746.833,3471.252)(6789.765,3452.469)(6800.498,3498.085)
\path(6540,2152)(6810,2692)
\path(6783.167,2571.252)(6810.000,2692.000)(6729.502,2598.085)
\path(6540,2152)(6990,3052) \path(6540,2152)(6090,3052)
\path(6540,2152)(6090,3052) \path(6540,2152)(6270,2692)
\path(6540,2152)(6270,2692)
\whiten\path(6350.498,2598.085)(6270.000,2692.000)(6296.833,2571.252)(6339.765,2552.469)(6350.498,2598.085)
\path(5640,2152)(5910,2692)
\path(5883.167,2571.252)(5910.000,2692.000)(5829.502,2598.085)
\path(5640,2152)(6090,3052) \path(5640,2152)(5190,3052)
\path(5640,2152)(5190,3052) \path(5640,2152)(5370,2692)
\path(5640,2152)(5370,2692)
\whiten\path(5450.498,2598.085)(5370.000,2692.000)(5396.833,2571.252)(5439.765,2552.469)(5450.498,2598.085)
\path(6090,1252)(6360,1792)
\path(6333.167,1671.252)(6360.000,1792.000)(6279.502,1698.085)
\path(6090,1252)(6540,2152) \path(6090,1252)(5640,2152)
\path(6090,1252)(5640,2152) \path(6090,1252)(5820,1792)
\path(6090,1252)(5820,1792)
\whiten\path(5900.498,1698.085)(5820.000,1792.000)(5846.833,1671.252)(5889.765,1652.469)(5900.498,1698.085)
\path(5640,3952)(5910,4492)
\path(5883.167,4371.252)(5910.000,4492.000)(5829.502,4398.085)
\path(5640,3952)(6090,4852) \path(6540,3952)(6090,4852)
\path(6540,3952)(6090,4852) \path(6540,3952)(6270,4492)
\path(6540,3952)(6270,4492)
\whiten\path(6350.498,4398.085)(6270.000,4492.000)(6296.833,4371.252)(6339.765,4352.469)(6350.498,4398.085)
\put(6090,4852){\blacken\ellipse{46}{46}}
\put(6090,4852){\ellipse{46}{46}}
\path(5640,3952)(5190,4852) \path(5640,3952)(5190,4852)
\path(5640,3952)(5370,4492) \path(5640,3952)(5370,4492)
\whiten\path(5450.498,4398.085)(5370.000,4492.000)(5396.833,4371.252)(5439.765,4352.469)(5450.498,4398.085)
\put(5025,4402){\makebox(0,0)[lb]{\smash{{\SetFigFont{8}{14.4}{\rmdefault}{\mddefault}{\updefault}$\gamma_1$}}}}
\path(6990,3052)(6990,4852) \path(6990,3052)(6990,4852)
\path(6990,3052)(6990,4087) \path(6990,3052)(6990,4087)
\whiten\path(7020.000,3967.000)(6990.000,4087.000)(6960.000,3967.000)(6990.000,4003.000)(7020.000,3967.000)
\put(7255,4357){\makebox(0,0)[lb]{\smash{{\SetFigFont{8}{14.4}{\rmdefault}{\mddefault}{\updefault}$\gamma'_1$}}}}
\put(4640,3502){\makebox(0,0)[lb]{\smash{{\SetFigFont{8}{14.4}{\rmdefault}{\mddefault}{\updefault}$\gamma_1$}}}}
\put(1815,6652){\blacken\ellipse{46}{46}}
\put(1815,6652){\ellipse{46}{46}}
\put(2265,5752){\blacken\ellipse{46}{46}}
\put(2265,5752){\ellipse{46}{46}}
\put(1365,5752){\blacken\ellipse{46}{46}}
\put(1365,5752){\ellipse{46}{46}}
\put(465,3952){\blacken\ellipse{46}{46}}
\put(465,3952){\ellipse{46}{46}}
\put(1365,3952){\blacken\ellipse{46}{46}}
\put(1365,3952){\ellipse{46}{46}}
\put(2265,3952){\blacken\ellipse{46}{46}}
\put(2265,3952){\ellipse{46}{46}}
\put(3165,3952){\blacken\ellipse{46}{46}}
\put(3165,3952){\ellipse{46}{46}}
\put(2715,3052){\blacken\ellipse{46}{46}}
\put(2715,3052){\ellipse{46}{46}}
\put(1815,3052){\blacken\ellipse{46}{46}}
\put(1815,3052){\ellipse{46}{46}}
\put(2265,2152){\blacken\ellipse{46}{46}}
\put(2265,2152){\ellipse{46}{46}}
\put(1815,1252){\blacken\ellipse{46}{46}}
\put(1815,1252){\ellipse{46}{46}}
\put(915,3052){\blacken\ellipse{46}{46}}
\put(915,3052){\ellipse{46}{46}}
\put(1365,2152){\blacken\ellipse{46}{46}}
\put(1365,2152){\ellipse{46}{46}}
\put(2715,4852){\blacken\ellipse{46}{46}}
\put(2715,4852){\ellipse{46}{46}}
\put(6090,6652){\blacken\ellipse{46}{46}}
\put(6090,6652){\ellipse{46}{46}}
\put(6540,5752){\blacken\ellipse{46}{46}}
\put(6540,5752){\ellipse{46}{46}}
\put(5640,5752){\blacken\ellipse{46}{46}}
\put(5640,5752){\ellipse{46}{46}}
\put(5640,3952){\blacken\ellipse{46}{46}}
\put(5640,3952){\ellipse{46}{46}}
\put(6540,3952){\blacken\ellipse{46}{46}}
\put(6540,3952){\ellipse{46}{46}}
\put(6990,3052){\blacken\ellipse{46}{46}}
\put(6990,3052){\ellipse{46}{46}}
\put(6090,3052){\blacken\ellipse{46}{46}}
\put(6090,3052){\ellipse{46}{46}}
\put(6540,2152){\blacken\ellipse{46}{46}}
\put(6540,2152){\ellipse{46}{46}}
\put(6090,1252){\blacken\ellipse{46}{46}}
\put(6090,1252){\ellipse{46}{46}}
\put(5190,3052){\blacken\ellipse{46}{46}}
\put(5190,3052){\ellipse{46}{46}}
\put(5640,2152){\blacken\ellipse{46}{46}}
\put(5640,2152){\ellipse{46}{46}}
\put(6990,4852){\blacken\ellipse{46}{46}}
\put(6990,4852){\ellipse{46}{46}}
\put(5190,4852){\blacken\ellipse{46}{46}}
\put(5190,4852){\ellipse{46}{46}} \path(915,4852)(645,5392)
\path(915,4852)(645,5392)
\whiten\path(725.498,5298.085)(645.000,5392.000)(671.833,5271.252)(725.498,5298.085)
\path(1365,5752)(1095,6292) \path(1365,5752)(1095,6292)
\whiten\path(1175.498,6198.085)(1095.000,6292.000)(1121.833,6171.252)(1175.498,6198.085)
\path(1815,6652)(1545,7192) \path(1815,6652)(1545,7192)
\whiten\path(1625.498,7098.085)(1545.000,7192.000)(1571.833,7071.252)(1625.498,7098.085)
\path(3165,3952)(3390,3502) \path(3165,3952)(3390,3502)
\path(3390,3502)(3345,3592) \path(3390,3502)(3345,3592)
\whiten\path(3425.498,3498.085)(3345.000,3592.000)(3371.833,3471.252)(3425.498,3498.085)
\path(2715,3052)(2940,2602) \path(2715,3052)(2940,2602)
\path(2940,2602)(2895,2692) \path(2940,2602)(2895,2692)
\whiten\path(2975.498,2598.085)(2895.000,2692.000)(2921.833,2571.252)(2975.498,2598.085)
\path(2265,2152)(2490,1702) \path(2265,2152)(2490,1702)
\path(2490,1702)(2445,1792) \path(2490,1702)(2445,1792)
\whiten\path(2525.498,1698.085)(2445.000,1792.000)(2471.833,1671.252)(2525.498,1698.085)
\path(1815,1252)(2040,802) \path(1815,1252)(2040,802)
\path(2040,802)(1995,892) \path(2040,802)(1995,892)
\whiten\path(2075.498,798.085)(1995.000,892.000)(2021.833,771.252)(2075.498,798.085)
\thicklines \path(3705,3817)(4335,3817)
\blacken\path(4095.000,3757.000)(4335.000,3817.000)(4095.000,3877.000)(4167.000,3817.000)(4095.000,3757.000)
\thinlines \path(5190,4852)(4920,5392) \path(5190,4852)(4920,5392)
\whiten\path(5000.498,5298.085)(4920.000,5392.000)(4946.833,5271.252)(5000.498,5298.085)
\path(5640,5752)(5370,6292) \path(5640,5752)(5370,6292)
\whiten\path(5450.498,6198.085)(5370.000,6292.000)(5396.833,6171.252)(5450.498,6198.085)
\path(6090,6652)(5820,7192) \path(6090,6652)(5820,7192)
\whiten\path(5900.498,7098.085)(5820.000,7192.000)(5846.833,7071.252)(5900.498,7098.085)
\path(6990,3052)(7215,2602) \path(6990,3052)(7215,2602)
\path(7215,2602)(7170,2692) \path(7215,2602)(7170,2692)
\whiten\path(7250.498,2598.085)(7170.000,2692.000)(7196.833,2571.252)(7250.498,2598.085)
\path(6540,2152)(6765,1702) \path(6540,2152)(6765,1702)
\path(6765,1702)(6720,1792) \path(6765,1702)(6720,1792)
\whiten\path(6800.498,1698.085)(6720.000,1792.000)(6746.833,1671.252)(6800.498,1698.085)
\path(6090,1252)(6315,802) \path(6090,1252)(6315,802)
\path(6315,802)(6270,892) \path(6315,802)(6270,892)
\whiten\path(6350.498,798.085)(6270.000,892.000)(6296.833,771.252)(6350.498,798.085)
\path(5190,3052)(4920,3592) \path(5190,3052)(4920,3592)
\whiten\path(5000.498,3498.085)(4920.000,3592.000)(4946.833,3471.252)(5000.498,3498.085)
\path(5190,3052)(5190,4852) \path(5190,3052)(5190,4852)
\path(5190,3052)(5190,4087) \path(5190,3052)(5190,4087)
\whiten\path(5220.000,3967.000)(5190.000,4087.000)(5160.000,3967.000)(5190.000,4003.000)(5220.000,3967.000)
\path(7215,4402)(7170,4492) \path(7215,4402)(7170,4492)
\whiten\path(7250.498,4398.085)(7170.000,4492.000)(7196.833,4371.252)(7250.498,4398.085)
\path(6990,4852)(7175,4482) \path(6990,4852)(7175,4482)
\put(345,5302){\makebox(0,0)[lb]{\smash{{\SetFigFont{8}{14.4}{\rmdefault}{\mddefault}{\updefault}$\gamma_2$}}}}
\put(795,6202){\makebox(0,0)[lb]{\smash{{\SetFigFont{8}{14.4}{\rmdefault}{\mddefault}{\updefault}$\gamma_3$}}}}
\put(1245,7102){\makebox(0,0)[lb]{\smash{{\SetFigFont{8}{14.4}{\rmdefault}{\mddefault}{\updefault}$\gamma_4$}}}}
\put(1135,82){\makebox(0,0)[lb]{\smash{{\SetFigFont{14}{16.8}{\rmdefault}{\mddefault}{\updefault}fig.
8a}}}}
\put(5410,82){\makebox(0,0)[lb]{\smash{{\SetFigFont{14}{16.8}{\rmdefault}{\mddefault}{\updefault}fig.
8b}}}}
\put(4620,5302){\makebox(0,0)[lb]{\smash{{\SetFigFont{8}{14.4}{\rmdefault}{\mddefault}{\updefault}$\gamma_2$}}}}
\put(5070,6202){\makebox(0,0)[lb]{\smash{{\SetFigFont{8}{14.4}{\rmdefault}{\mddefault}{\updefault}$\gamma_3$}}}}
\put(5520,7102){\makebox(0,0)[lb]{\smash{{\SetFigFont{8}{14.4}{\rmdefault}{\mddefault}{\updefault}$\gamma_4$}}}}
\put(5510,3502){\makebox(0,0)[lb]{\smash{{\SetFigFont{8}{14.4}{\rmdefault}{\mddefault}{\updefault}$\gamma'_1$}}}}
\put(365,3502){\makebox(0,0)[lb]{\smash{{\SetFigFont{8}{14.4}{\rmdefault}{\mddefault}{\updefault}$\gamma'_1$}}}}
\put(785,2602){\makebox(0,0)[lb]{\smash{{\SetFigFont{8}{14.4}{\rmdefault}{\mddefault}{\updefault}$\gamma'_2$}}}}
\put(5060,2602){\makebox(0,0)[lb]{\smash{{\SetFigFont{8}{14.4}{\rmdefault}{\mddefault}{\updefault}$\gamma'_2$}}}}
\put(1235,1702){\makebox(0,0)[lb]{\smash{{\SetFigFont{8}{14.4}{\rmdefault}{\mddefault}{\updefault}$\gamma'_3$}}}}
\put(1655,802){\makebox(0,0)[lb]{\smash{{\SetFigFont{8}{14.4}{\rmdefault}{\mddefault}{\updefault}$\gamma'_4$}}}}
\put(5510,1702){\makebox(0,0)[lb]{\smash{{\SetFigFont{8}{14.4}{\rmdefault}{\mddefault}{\updefault}$\gamma'_3$}}}}
\put(5960,802){\makebox(0,0)[lb]{\smash{{\SetFigFont{8}{14.4}{\rmdefault}{\mddefault}{\updefault}$\gamma'_4$}}}}
\end{picture}
} \hspace{-10mm} \setlength{\unitlength}{0.0004in}
\begingroup\makeatletter\ifx\SetFigFont\undefined%
\gdef\SetFigFont#1#2#3#4#5{%
  \reset@font\fontsize{#1}{#2pt}%
  \fontfamily{#3}\fontseries{#4}\fontshape{#5}%
  \selectfont}%
\fi\endgroup%
{\renewcommand{\dashlinestretch}{30}
\begin{picture}(8360,7129)(0,-10)
\put(1867,6517){\blacken\ellipse{46}{46}}
\put(1867,6517){\ellipse{46}{46}}
\put(2317,5617){\blacken\ellipse{46}{46}}
\put(2317,5617){\ellipse{46}{46}}
\put(1417,5617){\blacken\ellipse{46}{46}}
\put(1417,5617){\ellipse{46}{46}}
\put(1417,3817){\blacken\ellipse{46}{46}}
\put(1417,3817){\ellipse{46}{46}}
\put(2317,3817){\blacken\ellipse{46}{46}}
\put(2317,3817){\ellipse{46}{46}}
\put(1867,2917){\blacken\ellipse{46}{46}}
\put(1867,2917){\ellipse{46}{46}}
\put(2317,2017){\blacken\ellipse{46}{46}}
\put(2317,2017){\ellipse{46}{46}}
\put(967,2917){\blacken\ellipse{46}{46}}
\put(967,2917){\ellipse{46}{46}}
\put(1417,2017){\blacken\ellipse{46}{46}}
\put(1417,2017){\ellipse{46}{46}}
\put(2767,4717){\blacken\ellipse{46}{46}}
\put(2767,4717){\ellipse{46}{46}}
\put(967,4717){\blacken\ellipse{46}{46}}
\put(967,4717){\ellipse{46}{46}}
\put(2767,2917){\blacken\ellipse{46}{46}}
\put(2767,2917){\ellipse{46}{46}}
\put(1867,1117){\blacken\ellipse{46}{46}}
\put(1867,1117){\ellipse{46}{46}}
\put(1867,4717){\blacken\ellipse{46}{46}}
\put(1867,4717){\ellipse{46}{46}}
\put(4702,6517){\blacken\ellipse{46}{46}}
\put(4702,6517){\ellipse{46}{46}}
\put(5152,5617){\blacken\ellipse{46}{46}}
\put(5152,5617){\ellipse{46}{46}}
\put(4252,5617){\blacken\ellipse{46}{46}}
\put(4252,5617){\ellipse{46}{46}}
\put(4702,1117){\blacken\ellipse{46}{46}}
\put(4702,1117){\ellipse{46}{46}}
\put(4252,2017){\blacken\ellipse{46}{46}}
\put(4252,2017){\ellipse{46}{46}}
\put(5152,2017){\blacken\ellipse{46}{46}}
\put(5152,2017){\ellipse{46}{46}}
\put(4252,3817){\blacken\ellipse{46}{46}}
\put(4252,3817){\ellipse{46}{46}}
\put(5152,3817){\blacken\ellipse{46}{46}}
\put(5152,3817){\ellipse{46}{46}}
\put(4702,2917){\blacken\ellipse{46}{46}}
\put(4702,2917){\ellipse{46}{46}}
\put(4702,4717){\blacken\ellipse{46}{46}}
\put(4702,4717){\ellipse{46}{46}}
\put(7267,6517){\blacken\ellipse{46}{46}}
\put(7267,6517){\ellipse{46}{46}}
\put(7267,2917){\blacken\ellipse{46}{46}}
\put(7267,2917){\ellipse{46}{46}}
\put(7267,1117){\blacken\ellipse{46}{46}}
\put(7267,1117){\ellipse{46}{46}}
\put(7267,4717){\blacken\ellipse{46}{46}}
\put(7267,4717){\ellipse{46}{46}}
\put(6817,3817){\blacken\ellipse{46}{46}}
\put(6817,3817){\ellipse{46}{46}}
\put(7717,3817){\blacken\ellipse{46}{46}}
\put(7717,3817){\ellipse{46}{46}} \path(967,4717)(697,5257)
\path(967,4717)(697,5257)
\whiten\path(777.498,5163.085)(697.000,5257.000)(723.833,5136.252)(777.498,5163.085)
\path(1417,5617)(1147,6157) \path(1417,5617)(1147,6157)
\whiten\path(1227.498,6063.085)(1147.000,6157.000)(1173.833,6036.252)(1227.498,6063.085)
\path(1867,6517)(1597,7057) \path(1867,6517)(1597,7057)
\whiten\path(1677.498,6963.085)(1597.000,7057.000)(1623.833,6936.252)(1677.498,6963.085)
\path(967,2917)(697,3457) \path(967,2917)(697,3457)
\whiten\path(777.498,3363.085)(697.000,3457.000)(723.833,3336.252)(777.498,3363.085)
\thicklines \path(22,3817)(652,3817)
\blacken\path(412.000,3757.000)(652.000,3817.000)(412.000,3877.000)(484.000,3817.000)(412.000,3757.000)
\path(5782,3817)(6412,3817)
\blacken\path(6172.000,3757.000)(6412.000,3817.000)(6172.000,3877.000)(6244.000,3817.000)(6172.000,3757.000)
\path(3082,3817)(3712,3817)
\blacken\path(3472.000,3757.000)(3712.000,3817.000)(3472.000,3877.000)(3544.000,3817.000)(3472.000,3757.000)
\thinlines \path(967,4717)(1237,5257)
\path(1210.167,5136.252)(1237.000,5257.000)(1156.502,5163.085)
\path(967,4717)(1417,5617) \path(1867,4717)(1417,5617)
\path(1867,4717)(1417,5617) \path(1867,4717)(1597,5257)
\path(1867,4717)(1597,5257)
\whiten\path(1677.498,5163.085)(1597.000,5257.000)(1623.833,5136.252)(1666.765,5117.469)(1677.498,5163.085)
\path(2767,4717)(2992,4267) \path(2767,4717)(2992,4267)
\path(2317,3817)(2587,4357)
\path(2560.167,4236.252)(2587.000,4357.000)(2506.502,4263.085)
\path(2317,3817)(2767,4717) \path(2767,4717)(2317,5617)
\path(2767,4717)(2317,5617) \path(2767,4717)(2497,5257)
\path(2767,4717)(2497,5257)
\whiten\path(2577.498,5163.085)(2497.000,5257.000)(2523.833,5136.252)(2566.765,5117.469)(2577.498,5163.085)
\path(967,2917)(1237,3457)
\path(1210.167,3336.252)(1237.000,3457.000)(1156.502,3363.085)
\path(967,2917)(1417,3817) \path(1417,3817)(1687,4357)
\path(1660.167,4236.252)(1687.000,4357.000)(1606.502,4263.085)
\path(1417,3817)(1867,4717) \path(2767,2917)(2992,2467)
\path(2767,2917)(2992,2467) \path(1867,2917)(1417,3817)
\path(1867,2917)(1417,3817) \path(1417,3817)(967,4717)
\path(1867,2917)(2137,3457)
\path(2110.167,3336.252)(2137.000,3457.000)(2056.502,3363.085)
\path(1867,2917)(2317,3817) \path(2317,2017)(2587,2557)
\path(2560.167,2436.252)(2587.000,2557.000)(2506.502,2463.085)
\path(2317,2017)(2767,2917) \path(2317,2017)(2542,1567)
\path(2317,2017)(2542,1567) \path(1417,2017)(1687,2557)
\path(1660.167,2436.252)(1687.000,2557.000)(1606.502,2463.085)
\path(1417,2017)(1867,2917) \path(1867,1117)(2137,1657)
\path(2110.167,1536.252)(2137.000,1657.000)(2056.502,1563.085)
\path(1867,1117)(2317,2017) \path(1867,1117)(2092,667)
\path(1867,1117)(2092,667) \path(1867,1117)(1417,2017)
\path(1867,1117)(1417,2017) \path(4252,5617)(3982,6157)
\path(4252,5617)(3982,6157)
\whiten\path(4062.498,6063.085)(3982.000,6157.000)(4008.833,6036.252)(4062.498,6063.085)
\path(4702,6517)(4432,7057) \path(4702,6517)(4432,7057)
\whiten\path(4512.498,6963.085)(4432.000,7057.000)(4458.833,6936.252)(4512.498,6963.085)
\path(4252,2017)(3982,2557) \path(4252,2017)(3982,2557)
\whiten\path(4062.498,2463.085)(3982.000,2557.000)(4008.833,2436.252)(4062.498,2463.085)
\path(4702,1117)(4972,1657)
\path(4945.167,1536.252)(4972.000,1657.000)(4891.502,1563.085)
\path(4702,1117)(5152,2017) \path(4702,1117)(4927,667)
\path(4702,1117)(4927,667) \path(5152,2017)(4702,2917)
\path(5152,2017)(4702,2917) \path(5152,2017)(4882,2557)
\path(5152,2017)(4882,2557)
\whiten\path(4962.498,2463.085)(4882.000,2557.000)(4908.833,2436.252)(4951.765,2417.469)(4962.498,2463.085)
\path(5152,2017)(5377,1567) \path(5152,2017)(5377,1567)
\path(4702,1117)(4252,2017) \path(4702,1117)(4252,2017)
\path(4702,1117)(4432,1657) \path(4702,1117)(4432,1657)
\whiten\path(4512.498,1563.085)(4432.000,1657.000)(4458.833,1536.252)(4501.765,1517.469)(4512.498,1563.085)
\path(4252,2017)(4522,2557)
\path(4495.167,2436.252)(4522.000,2557.000)(4441.502,2463.085)
\path(4252,2017)(4702,2917) \path(4702,2917)(4252,3817)
\path(4702,2917)(4252,3817) \path(4702,2917)(4432,3457)
\path(4702,2917)(4432,3457)
\whiten\path(4512.498,3363.085)(4432.000,3457.000)(4458.833,3336.252)(4501.765,3317.469)(4512.498,3363.085)
\path(5152,3817)(5377,3367) \path(5152,3817)(5377,3367)
\path(4702,2917)(4972,3457)
\path(4945.167,3336.252)(4972.000,3457.000)(4891.502,3363.085)
\path(4702,2917)(5152,3817) \path(4252,3817)(3982,4357)
\path(4252,3817)(3982,4357)
\whiten\path(4062.498,4263.085)(3982.000,4357.000)(4008.833,4236.252)(4062.498,4263.085)
\path(5152,5617)(4702,6517) \path(5152,5617)(4702,6517)
\path(5152,5617)(4882,6157) \path(5152,5617)(4882,6157)
\whiten\path(4962.498,6063.085)(4882.000,6157.000)(4908.833,6036.252)(4951.765,6017.469)(4962.498,6063.085)
\path(4252,5617)(4522,6157)
\path(4495.167,6036.252)(4522.000,6157.000)(4441.502,6063.085)
\path(4252,5617)(4702,6517) \path(5152,5617)(5377,5167)
\path(5152,5617)(5377,5167) \path(4702,4717)(4252,5617)
\path(4702,4717)(4252,5617) \path(4702,4717)(4432,5257)
\path(4702,4717)(4432,5257)
\whiten\path(4512.498,5163.085)(4432.000,5257.000)(4458.833,5136.252)(4501.765,5117.469)(4512.498,5163.085)
\path(4702,4717)(4972,5257)
\path(4945.167,5136.252)(4972.000,5257.000)(4891.502,5163.085)
\path(4702,4717)(5152,5617) \path(4252,3817)(4522,4357)
\path(4495.167,4236.252)(4522.000,4357.000)(4441.502,4263.085)
\path(4252,3817)(4702,4717) \path(5152,3817)(4702,4717)
\path(5152,3817)(4702,4717) \path(5152,3817)(4882,4357)
\path(5152,3817)(4882,4357)
\whiten\path(4962.498,4263.085)(4882.000,4357.000)(4908.833,4236.252)(4951.765,4217.469)(4962.498,4263.085)
\path(7267,6517)(6997,7057) \path(7267,6517)(6997,7057)
\whiten\path(7077.498,6963.085)(6997.000,7057.000)(7023.833,6936.252)(7077.498,6963.085)
\path(7267,1117)(6997,1657) \path(7267,1117)(6997,1657)
\whiten\path(7077.498,1563.085)(6997.000,1657.000)(7023.833,1536.252)(7077.498,1563.085)
\path(7267,4717)(6997,5257) \path(7267,4717)(6997,5257)
\whiten\path(7077.498,5163.085)(6997.000,5257.000)(7023.833,5136.252)(7077.498,5163.085)
\path(7267,6517)(7492,6067) \path(7267,6517)(7492,6067)
\path(7717,3817)(7267,4717) \path(7717,3817)(7267,4717)
\path(7717,3817)(7447,4357) \path(7717,3817)(7447,4357)
\whiten\path(7527.498,4263.085)(7447.000,4357.000)(7473.833,4236.252)(7516.765,4217.469)(7527.498,4263.085)
\path(6817,3817)(7087,4357)
\path(7060.167,4236.252)(7087.000,4357.000)(7006.502,4263.085)
\path(6817,3817)(7267,4717) \path(6817,3817)(6547,4357)
\path(6817,3817)(6547,4357)
\whiten\path(6627.498,4263.085)(6547.000,4357.000)(6573.833,4236.252)(6627.498,4263.085)
\path(7267,2917)(7537,3457)
\path(7510.167,3336.252)(7537.000,3457.000)(7456.502,3363.085)
\path(7267,2917)(7717,3817) \path(7267,2917)(6817,3817)
\path(7267,2917)(6817,3817) \path(7267,2917)(6997,3457)
\path(7267,2917)(6997,3457)
\whiten\path(7077.498,3363.085)(6997.000,3457.000)(7023.833,3336.252)(7066.765,3317.469)(7077.498,3363.085)
\path(7717,3817)(7942,3367) \path(7717,3817)(7942,3367)
\path(7267,2917)(7492,2467) \path(7267,2917)(7492,2467)
\path(7267,1117)(7492,667) \path(7267,1117)(7492,667)
\path(1867,4717)(2137,5257)
\path(2110.167,5136.252)(2137.000,5257.000)(2056.502,5163.085)
\path(1867,4717)(2317,5617) \path(2317,5617)(1867,6517)
\path(2317,5617)(1867,6517) \path(2317,5617)(2047,6157)
\path(2317,5617)(2047,6157)
\whiten\path(2127.498,6063.085)(2047.000,6157.000)(2073.833,6036.252)(2116.765,6017.469)(2127.498,6063.085)
\path(1417,5617)(1687,6157)
\path(1660.167,6036.252)(1687.000,6157.000)(1606.502,6063.085)
\path(1417,5617)(1867,6517) \path(2317,3817)(1867,4717)
\path(2317,3817)(1867,4717) \path(2317,3817)(2047,4357)
\path(2317,3817)(2047,4357)
\whiten\path(2127.498,4263.085)(2047.000,4357.000)(2073.833,4236.252)(2116.765,4217.469)(2127.498,4263.085)
\put(1097,82){\makebox(0,0)[lb]{\smash{{\SetFigFont{14}{16.8}{\rmdefault}{\mddefault}{\updefault}fig.
8c}}}}
\put(842,4267){\makebox(0,0)[lb]{\smash{{\SetFigFont{8}{12.0}{\rmdefault}{\mddefault}{\updefault}$\gamma'_1$}}}}
\put(387,3367){\makebox(0,0)[lb]{\smash{{\SetFigFont{8}{12.0}{\rmdefault}{\mddefault}{\updefault}$\gamma_1$}}}}
\put(3932,82){\makebox(0,0)[lb]{\smash{{\SetFigFont{14}{16.8}{\rmdefault}{\mddefault}{\updefault}fig.
8d}}}}
\put(6452,82){\makebox(0,0)[lb]{\smash{{\SetFigFont{14}{16.8}{\rmdefault}{\mddefault}{\updefault}fig.
8e}}}}
\put(287,5167){\makebox(0,0)[lb]{\smash{{\SetFigFont{8}{12.0}{\rmdefault}{\mddefault}{\updefault}$\gamma_2$}}}}
\put(837,2467){\makebox(0,0)[lb]{\smash{{\SetFigFont{8}{12.0}{\rmdefault}{\mddefault}{\updefault}$\gamma'_2$}}}}
\put(1287,1567){\makebox(0,0)[lb]{\smash{{\SetFigFont{8}{12.0}{\rmdefault}{\mddefault}{\updefault}$\gamma'_3$}}}}
\put(1737,667){\makebox(0,0)[lb]{\smash{{\SetFigFont{8}{12.0}{\rmdefault}{\mddefault}{\updefault}$\gamma'_4$}}}}
\put(837,6067){\makebox(0,0)[lb]{\smash{{\SetFigFont{8}{12.0}{\rmdefault}{\mddefault}{\updefault}$\gamma_3$}}}}
\put(1287,6967){\makebox(0,0)[lb]{\smash{{\SetFigFont{8}{12.0}{\rmdefault}{\mddefault}{\updefault}$\gamma_4$}}}}
\put(4122,5167){\makebox(0,0)[lb]{\smash{{\SetFigFont{8}{12.0}{\rmdefault}{\mddefault}{\updefault}$\gamma'_1$}}}}
\put(3672,6067){\makebox(0,0)[lb]{\smash{{\SetFigFont{8}{12.0}{\rmdefault}{\mddefault}{\updefault}$\gamma_3$}}}}
\put(3672,4267){\makebox(0,0)[lb]{\smash{{\SetFigFont{8}{12.0}{\rmdefault}{\mddefault}{\updefault}$\gamma_2$}}}}
\put(4122,3367){\makebox(0,0)[lb]{\smash{{\SetFigFont{8}{12.0}{\rmdefault}{\mddefault}{\updefault}$\gamma'_2$}}}}
\put(3672,2467){\makebox(0,0)[lb]{\smash{{\SetFigFont{8}{12.0}{\rmdefault}{\mddefault}{\updefault}$\gamma_1$}}}}
\put(4122,1567){\makebox(0,0)[lb]{\smash{{\SetFigFont{8}{12.0}{\rmdefault}{\mddefault}{\updefault}$\gamma'_3$}}}}
\put(4572,667){\makebox(0,0)[lb]{\smash{{\SetFigFont{8}{12.0}{\rmdefault}{\mddefault}{\updefault}$\gamma'_4$}}}}
\put(6687,1567){\makebox(0,0)[lb]{\smash{{\SetFigFont{8}{12.0}{\rmdefault}{\mddefault}{\updefault}$\gamma_1$}}}}
\put(7137,667){\makebox(0,0)[lb]{\smash{{\SetFigFont{8}{12.0}{\rmdefault}{\mddefault}{\updefault}$\gamma'_4$}}}}
\put(6687,3367){\makebox(0,0)[lb]{\smash{{\SetFigFont{8}{12.0}{\rmdefault}{\mddefault}{\updefault}$\gamma'_2$}}}}
\put(7137,2467){\makebox(0,0)[lb]{\smash{{\SetFigFont{8}{12.0}{\rmdefault}{\mddefault}{\updefault}$\gamma'_3$}}}}
\put(6237,4267){\makebox(0,0)[lb]{\smash{{\SetFigFont{8}{12.0}{\rmdefault}{\mddefault}{\updefault}$\gamma_2$}}}}
\put(6687,5167){\makebox(0,0)[lb]{\smash{{\SetFigFont{8}{12.0}{\rmdefault}{\mddefault}{\updefault}$\gamma_3$}}}}
\put(7137,6067){\makebox(0,0)[lb]{\smash{{\SetFigFont{8}{12.0}{\rmdefault}{\mddefault}{\updefault}$\gamma'_1$}}}}
\put(6687,6967){\makebox(0,0)[lb]{\smash{{\SetFigFont{8}{12.0}{\rmdefault}{\mddefault}{\updefault}$\gamma_4$}}}}
\put(4122,6967){\makebox(0,0)[lb]{\smash{{\SetFigFont{8}{12.0}{\rmdefault}{\mddefault}{\updefault}$\gamma_4$}}}}
\path(2992,4267)(2947,4357) \path(2992,4267)(2947,4357)
\whiten\path(3027.498,4263.085)(2947.000,4357.000)(2973.833,4236.252)(3027.498,4263.085)
\path(2992,2467)(2947,2557) \path(2992,2467)(2947,2557)
\whiten\path(3027.498,2463.085)(2947.000,2557.000)(2973.833,2436.252)(3027.498,2463.085)
\path(1867,2917)(1597,3457) \path(1867,2917)(1597,3457)
\whiten\path(1677.498,3363.085)(1597.000,3457.000)(1623.833,3336.252)(1666.765,3317.469)(1677.498,3363.085)
\path(1417,3817)(1147,4357) \path(1417,3817)(1147,4357)
\whiten\path(1227.498,4263.085)(1147.000,4357.000)(1173.833,4236.252)(1216.765,4217.469)(1227.498,4263.085)
\path(2767,2917)(2317,3817) \path(2767,2917)(2317,3817)
\path(2542,1567)(2497,1657) \path(2542,1567)(2497,1657)
\whiten\path(2577.498,1563.085)(2497.000,1657.000)(2523.833,1536.252)(2577.498,1563.085)
\path(2092,667)(2047,757) \path(2092,667)(2047,757)
\whiten\path(2127.498,663.085)(2047.000,757.000)(2073.833,636.252)(2127.498,663.085)
\path(1417,2017)(967,2917) \path(1417,2017)(967,2917)
\path(2317,2017)(1867,2917) \path(2317,2017)(1867,2917)
\path(2767,2917)(2497,3457) \path(2767,2917)(2497,3457)
\whiten\path(2577.498,3363.085)(2497.000,3457.000)(2523.833,3336.252)(2566.765,3317.469)(2577.498,3363.085)
\path(2317,2017)(2047,2557) \path(2317,2017)(2047,2557)
\whiten\path(2127.498,2463.085)(2047.000,2557.000)(2073.833,2436.252)(2116.765,2417.469)(2127.498,2463.085)
\path(1417,2017)(1147,2557) \path(1417,2017)(1147,2557)
\whiten\path(1227.498,2463.085)(1147.000,2557.000)(1173.833,2436.252)(1216.765,2417.469)(1227.498,2463.085)
\path(1867,1117)(1597,1657) \path(1867,1117)(1597,1657)
\whiten\path(1677.498,1563.085)(1597.000,1657.000)(1623.833,1536.252)(1666.765,1517.469)(1677.498,1563.085)
\path(4927,667)(4882,757) \path(4927,667)(4882,757)
\whiten\path(4962.498,663.085)(4882.000,757.000)(4908.833,636.252)(4962.498,663.085)
\path(5377,1567)(5332,1657) \path(5377,1567)(5332,1657)
\whiten\path(5412.498,1563.085)(5332.000,1657.000)(5358.833,1536.252)(5412.498,1563.085)
\path(5377,3367)(5332,3457) \path(5377,3367)(5332,3457)
\whiten\path(5412.498,3363.085)(5332.000,3457.000)(5358.833,3336.252)(5412.498,3363.085)
\path(5377,5167)(5332,5257) \path(5377,5167)(5332,5257)
\whiten\path(5412.498,5163.085)(5332.000,5257.000)(5358.833,5136.252)(5412.498,5163.085)
\path(7492,6067)(7447,6157) \path(7492,6067)(7447,6157)
\whiten\path(7527.498,6063.085)(7447.000,6157.000)(7473.833,6036.252)(7527.498,6063.085)
\path(7942,3367)(7897,3457) \path(7942,3367)(7897,3457)
\whiten\path(7977.498,3363.085)(7897.000,3457.000)(7923.833,3336.252)(7977.498,3363.085)
\path(7492,2467)(7447,2557) \path(7492,2467)(7447,2557)
\whiten\path(7527.498,2463.085)(7447.000,2557.000)(7473.833,2436.252)(7527.498,2463.085)
\path(7492,667)(7447,757) \path(7492,667)(7447,757)
\whiten\path(7527.498,663.085)(7447.000,757.000)(7473.833,636.252)(7527.498,663.085)
\end{picture}
}


The support of distribution $\prod\limits_{j=1}^N\delta(\gamma'_j-\gamma_{N+1-j})$ belongs to the domain $|\gamma'_j-\gamma_{N+1-j}|<\epsilon$, $j=1,\ldots,N$, in which $\phi(\gamma';\gamma)=1$ due to the property~\ref{prop3}. Substituting $\gamma'_j=\gamma_{N+1-j}$ in the product of Gamma-functions in right hand side of~\eqref{int_psi} we derive~\eqref{I_delta}
with the expression~\eqref{measure} for the Sklyanin measure. \qed \\

\section{Properties of the $\Lambda$-operators}
\label{Pr_Lambda}

Baxter's $Q$-operators $\Hat Q_N(u)$ described in~\cite{Pasquier}
for the periodic Toda chain satisfy the following properties.
\begin{itemize}
\item[(a)] These operators commute for different values of the spectral parameters: \\
$[\hat Q_N(u),\hat Q_N(v)]=0$.
\item[(b)] They commute with the transfer matrix $\hat
t_N(u)\hm=A_N(u)\hm+D_N(u)$ of the periodic Toda chain model:
$[\hat Q_N(u),\hat t_N(v)]=0$.
\item[(c)] $Q$-operator satisfies the Baxter equation
\begin{equation*} \label{Baxter_od_}
  \hat t_N(u)\hat Q_N(u)=i^N \hat Q_N(u+i\hbar)+i^{-N} \hat Q_N(u-i\hbar).
\end{equation*}
\end{itemize}

In this section the similar properties for the operators
$\Lambda_N(u)$ defined in section~\ref{Eig_OTCh} will be
established.

\begin{prop}
 $\Lambda$-operator has the following properties:
\begin{itemize}
\item[{\scshape (i)}] $\Lambda_N(u)\Lambda_{N-1}(v)=\Lambda_N(v)\Lambda_{N-1}(u),$

\item[{\scshape (ii)}] $A_N(u)\Lambda_N(v)=(u-v)\Lambda_N(v)A_{N-1}(u),$

\item[{\scshape (iii)}] $C_N(u)\Lambda_N(u)=i^{-N-1}\Lambda_N(u-i\hbar),$

\item[{\scshape (iv)}] $B_N(u)\Lambda_N(u)=i^{N-1}\Lambda_N(u+i\hbar).$
\end{itemize}
\end{prop}

{\bfseries Proof.} The equality \textsc{(i)} was proved in the
theorem~\ref{Th_psi}.
Property \textsc{(ii)} have been proved in \cite{Kharchev_GG} by
the direct calculation. Below we give a simplified version of this
property. Let us
notice that \textsc{(ii)} is valid on arbitrary $N-1$-particle eigenfunction
$\psi_{\gamma_1,\ldots,\gamma_{N-1}}(y_1,\ldots, y_{N-1})$.
Indeed, in according with the equation~\eqref{psi_def} the left hand
side of \textsc{(ii)} is
\begin{multline} \label{lhs_pr_ii}
A_N(u)(\Lambda_N(v)\psi_{\gamma_1,\ldots,\gamma_{N-1}})(x_1,\ldots, x_N)=
A_N(u)\psi_{\gamma_1,\ldots,\gamma_{N-1},v}(x_1,\ldots, x_N)=\\
 =(u-v)\prod_{k=1}^{N-1}(u-\gamma_k)\psi_{\gamma_1,\ldots,\gamma_{N-1},v}(x_1,\ldots, x_N)
\end{multline}
while the right hand side is
\begin{multline} \label{rhs_pr_ii}
(u-v)(\Lambda_N(v)A_{N-1}(u)\psi_{\gamma_1,\ldots,\gamma_{N-1}})(x_1,\ldots,x_N)=\\
 =(u-v)\prod_{k=1}^{N-1}(u-\gamma_k)(\Lambda_N(v)\psi_{\gamma_1,\ldots,\gamma_{N-1}})(x_1,\ldots, x_N)=\\
  =(u-v)\prod_{k=1}^{N-1}(u-\gamma_k)\psi_{\gamma_1,\ldots,\gamma_{N-1},v}(x_1,\ldots, x_N).
\end{multline}
Comparing~\eqref{lhs_pr_ii} and \eqref{rhs_pr_ii} one concludes
that the equality~\textsc{(ii)} is valid on the functions
$\psi_{\gamma}(x)$. Due to the completeness of the system of these
functions any function that belongs to the domain of definition of
the operators $A_{N-1}(u)$ and $\Lambda_N(u)$ can be represented
in the form~\eqref{f_psi}. Therefore the property \textsc{(ii)} is
valid on any function of $N-1$ variables where the action of the
operators $A_{N-1}(u)$ and $\Lambda_N(v)$ is well defined.  \\

To prove \textsc{(iii)} we need the formula for an action of the
operators $A_m(u)$ on the kernel of the operator $\Lambda_N(u)$:
\begin{equation} \label{AmLam}
 A_m(u)\Lambda_{u}(x_1,\ldots,x_N;y_1,\ldots,y_{N-1})=
 (-i)^{m}e^{\sum\limits_{j=1}^m(x_j-y_j)}\Lambda_{u}(x_1,\ldots,x_N;y_1,\ldots,y_{N-1})
\end{equation}
for $m=0,\ldots,N-1$, ($A_0(u)\hm=1$ is implied). In the case
$m\hm=0$ and $m\hm=1$ this equality is obvious. For
$m=2,\ldots,N-1$ it can be proved by induction using the
relation~\eqref{AArec}. Indeed
\begin{multline} \label{iii_proof_ind}
 A_m(u)\Lambda_{u}(x;y)=(u+i\hbar\partial_{x_m})A_{m-1}(u)\Lambda_{u}(x;y)-e^{x_{m-1}-x_m}A_{m-2}(u)\Lambda_{u}(x;y)=\\
 =\bigl[(-i)^{m}e^{\sum\limits_{j=1}^{m-1}(x_j-y_j)}(e^{x_m-y_m}-e^{y_{m-1}-x_m})-(-i)^{m-2}e^{x_{m-1}-x_m}e^{\sum\limits_{j=1}^{m-2}(x_j-y_j)}\bigr]\Lambda_{u}(x;y)=\\
 =(-i)^{m}e^{\sum\limits_{j=1}^m(x_j-y_j)}\Lambda_{u}(x;y).
\end{multline}
The formula~\eqref{AmLam} for $m=N-1$ and the
relation~\eqref{Crec} give
\begin{multline*}
 C_N(u)\Lambda_{u}(x;y)=-e^{x_N}A_{N-1}(u)\Lambda_{u}(x;y)= \\
  =-(-i)^{N-1}e^{x_N+\sum_{j=1}^{N-1}(x_j-y_j)}\Lambda_{u}(x;y)=(-i)^{N+1}\Lambda_{u-i\hbar}(x;y),
\end{multline*}
what means in turn the equality \textsc{(iii)}. \\

Note that if one sets  $m=N$ in~\eqref{iii_proof_ind} then the
term containing $e^{x_m-y_m}$ does not arise and, consequently, we
obtain zero in the right hand side. Thus we have proved directly
that the kernel $\Lambda_{u}(x;y)$ satisfies to the
equation~\eqref{Apsi0}. \\

Analogously, to prove \textsc{(iv)} we used
\begin{equation} \label{BBrec}
 B_N(u)=(u-p_N)B_{N-1}(u)-e^{x_{N-1}-x_N}B_{N-2}(u),
\end{equation}
and the action of operator $B_m(u)$ on the kernel of the operator
$\Lambda_N(u)$
\begin{equation} \label{BmLam}
 B_m(u)\Lambda_{u}(x;y)=e^{-x_1}(-i)^{m-1}\sum_{k=1}^m(-1)^{k+1}\prod_{j=2}^k e^{y_{j-1}-x_j}
   \prod_{s=k+1}^m e^{x_s-y_s}\Lambda_{u}(x;y)
\end{equation}
which can be also proved by induction.
Exploring the relation~\eqref{BBrec} and the formula~\eqref{BmLam}
for $m=N-1, N-2$ we obtain
\begin{multline*}
 B_{N}(u)\Lambda_{u}(x;y)=e^{-x_1}(-i)^{N-1}\sum_{k=1}^{N-1} (-1)^{k+1}\prod_{j=2}^k e^{y_{j-1}-x_j}
    \prod_{s=k+1}^{N-1} e^{x_s-y_s}(-e^{y_{N-1}-x_N})\Lambda_{u}(x;y)-\\
  -e^{-x_1}(-i)^{N-3}\sum_{k=1}^{N-2} (-1)^{k+1}\prod_{j=2}^k e^{y_{j-1}-x_j}
    \prod_{s=k+1}^{N-2} e^{x_s-y_s} e^{x_{N-1}-x_N}\Lambda_{u}(x;y)=\\
  =-e^{-x_1}(-i)^{N-1}(-1)^N \prod_{j=2}^N e^{y_{j-1}-x_j}\Lambda_{u}(x;y)=i^{N-1}\Lambda_{u+i\hbar}(x;y).
\end{multline*}

Let us note that the properties~\textsc{(iii)}, \textsc{(iv)} up to
the factors $(-i)^{\mp N-1}$ can be derived from \textsc{(ii)}
using R-matrix formalism (see for example~\cite{Kharchev_OP},
\cite{Kharchev_O}, \cite{Kharchev_P}).
Thes properties lead to the Baxter equation for operator $\Lambda_{N-1}(u)$. Indeed, taking into account the recurrent relations~\eqref{Arec} and $D_N(u)=-e^{x_N}B_N(u)$, property~\textsc{(ii)} for $u=v$ and applying the properties~\textsc{(iii)}, \textsc{(iv)} one yields
\begin{align}
 \hat t_N(u)\Lambda_{N-1}(u)=i^{-N}e^{-x_N}\Lambda_{N-1}(u-i\hbar)+i^Ne^{x_N}\Lambda_{N-1}(u+i\hbar). \label{BE_Lamb}
\end{align}

\section{Conclusion}
\label{Conclusion}

The derivation of the eigenfunction of the open Toda chain can be considered as the fist step of the Separation of Variables problem for the periodic toda chain. As it was mentioned in the introduction the Separation of Variables is achieved by a special choice of the transition function. In the case of Toda chain this function should be chosen as follows~\cite{Kharchev_P}:
\begin{equation} \label{UEq}
 U_{\varepsilon,\gamma_1,\ldots,\gamma_{N-1}}(x_1,\ldots,x_N)=
  e^{\frac{i}{\hbar}\bigl(\varepsilon-\sum\limits_{j=1}^{N-1}\gamma_j\bigr)x_N}
   \psi_{\gamma_1,\ldots,\gamma_{N-1}}(x_1,\ldots,x_{N-1}).
\end{equation}
The eigenfunctions of the integrals of motion of the periodic Toda chain, which are the expansion coefficients for the operator $\hat t_N(u)$, are represented in the form
\begin{align*}
 \Psi_E(x_1,\ldots,x_N)=\int\limits_{\mathbb R^{N-1}} U_{E_1,\gamma_1,\ldots,\gamma_{N-1}}(x_1,\ldots,x_N) \Phi_E(\gamma_1,\ldots,\gamma_{N-1})\mu(\gamma)d\gamma.
\end{align*}

 The function~\eqref{UEq} can be calculated by induction: having an expression for the $(N-1)$-particle transition function one can yield the expression for the $N$-particle one. This method was proposed in the work~\cite{Kharchev_O}.  The authors of~\cite{Kharchev_O} obtain the recurrent formula integrating over the variables $\gamma_j$:
\begin{equation}  \label{psi_rec_gamma}
 \begin{split}
 \psi_{\lambda_1,\ldots,\lambda_N}&(x_1,\ldots,x_N)= \\
   &=\int\limits_{\mathbb R^{N-1}}
  e^{\frac{i}{\hbar}\bigl(\sum\limits_{k=1}^N\lambda_k-\sum\limits_{j=1}^{N-1}\gamma_j\bigr)x_N}
  \psi_{\gamma_1,\ldots,\gamma_{N-1}}(x_1,\ldots,x_{N-1})
  K(\lambda;\gamma) \mu(\gamma)d\gamma,
 \end{split}
\end{equation}
where $\mu(\gamma)$ is a Sklyanin measure described in section~\ref{Int_meas} and $K(\lambda;\gamma)$ is some kernel. This formula leads to the Mellin-Barns representation for the transition function. \\

In the present paper it is shown that the recurrent integration can be realised in terms of the coordinates $x_n$ \big(eq.~\eqref{psi_rec}\big). This leads in turn to the Gauss-Givental representation~\eqref{psi_GG}, which was obtained in~\cite{Givental}, \cite{Kharchev_GG} from the other circle of ideas. The integration over the coordinates in the formula~\eqref{psi_rec} implies actually an action of the operator $\Lambda_N(\gamma_N)$ on the $N$-particle eigenfunction. The function~\eqref{UEq} can be rewritten then in terms of the product of $\Lambda$-operators like~\eqref{psi}. This results in the fact that the function~\eqref{UEq} inherits the properties of the $\Lambda$-operators discussed in section~\ref{Pr_Lambda}. For example, the Weyl-invariance of this function is encoded in the property \textsc{(i)}. Due to the properties \textsc{(iii)} and \textsc{(iv)} the equation~\eqref{BE_Lamb} holds, and therefore the transition function~\eqref{UEq} satisfy the Baxter equation
\begin{align*}
 \hat t_N(\gamma_k)U_{\varepsilon,\gamma}(x)=i^{-N}U_{\varepsilon,\gamma-i\hbar\delta_k}(x)+i^N U_{\varepsilon,\gamma+i\hbar\delta_k}(x),
\end{align*}
where $\delta_k=(0,\ldots,1,\ldots,0)$ is a $k$-th base vector. Since the Sklyanin measure $\mu(\gamma)$ has necessary translational properties the eigenvalue equation
\begin{align*}
 \hat t_N(u)\Psi_E(x)=t_N(u;E)\Psi_E(x),
\end{align*}
where
\begin{align*}
 t_N(u;E)=\sum_{k=0}^N(-1)^k u^{N-k}E_k,
\end{align*}
is equivalent to the Baxter equation for the functions $\Phi_{\varepsilon}(\gamma)$
\begin{align}
 t_N(\gamma_k;E)\Phi_{\varepsilon}(\gamma)=i^N\Phi_{\varepsilon}(\gamma+i\hbar\delta_k)+i^{-N} \Phi_{\varepsilon}(\gamma-i\hbar\delta_k).
\end{align}
This leads in turn to the occasion to present these function in the form
\begin{align*}
 \Phi_\varepsilon(\gamma_1,\ldots,\gamma_{N-1})=\prod\limits_{j=1}^{N-1}c_\varepsilon(\gamma_j),
\end{align*}
that is to the separation of variables (see details in~\cite{Kharchev_P}). \\

Availability of two kind of recurrent formulae -- of type~\eqref{psi_rec_gamma} and of type~\eqref{psi_rec} -- is explained by the fact that the function $\psi_\gamma(x)$ can be regarded as well as a function of $x_n$ satisfying the differential equations~\eqref{psi_def} and in other hand as a function of $\gamma_j$ satisfying difference equations in $\gamma_j$ \cite{Babelon}, i.e. as a wave function of some dual model. The duality of the same kind appears in the Representation Theory. The infinite-dimensional Gelfand-Zetlin representation of Lie algebra $\mathfrak{gl}(N)$ by shift operators in $\gamma_j$ allows to obtain the Mellin-Barns integral representation~\cite{Kharchev_GZ}, while the Gauss representation of the same Lie algebra by differential operators in $x_n$ leads to Gauss-Givental representation~\cite{Kharchev_GG}. \\

We hope the method proposed for the $XXX$-model in~\cite{Derkachov} and developed here for the Toda chain (including the use of diagram technique) can be applied to other more complicated integrable systems.



\section*{Acknowledgments}

This work is a part of PhD which the author is preparing at Bogoliubov Laboratory of Theoretical Physics (JINR, Dubna, Russia) and at LAREMA (Univ.\,of Angers, France). He thanks both laboratories for excellent work conditions. He is thankful also to the French-Russian Network in Theoretical Physics and to prof. J.-M.\,Maillet for financial support for this joint PhD programme.\\

Author thanks to S.\,M.\,Kharchev, S.\,Z.\,Pakuliak and V.\,N.\,Rubtsov for useful advices and remarks.

\end{document}